\documentclass[aps,amsmath,twocolumn,amssymb,floatfix,superscriptaddress,nofootinbib,longbibliography]{revtex4}
\usepackage{mathtools}
\usepackage{braket}
\usepackage[dvipsnames]{xcolor}
\usepackage{float}
\usepackage{subfigure}
\usepackage{graphics}
\usepackage{bm}
\usepackage{amsmath,amssymb,bm}

\usepackage{braket}
\usepackage{float}
\usepackage{tikz}
\usepackage{blkarray}
\usepackage{pgffor}
\usepackage{float}
\usepackage{silence}
\usepackage[colorlinks=true,linktoc=page,linkcolor=BrickRed,citecolor=magenta,urlcolor=purple]{hyperref}

\mathchardef\mhyphen="2D 

\newcommand{\etal}{{\it et al.}}

\newcommand\bea{\begin{eqnarray}}
\newcommand\eea{\end{eqnarray}}
\newcommand\beq{\begin{equation}}  
\newcommand\eeq{\end{equation}}

\usepackage[normalem]{ulem}
\definecolor{lime}{HTML}{A6CE39}
\usepackage{sidecap,tikz}
\DeclareRobustCommand{\orcidicon}{\hspace{-1.0mm}
	\begin{tikzpicture}
		\draw[lime, fill=lime] (0.0,0.0) 
		circle [radius=0.15] 
		node[white] {{\fontfamily{qag}\selectfont \tiny \,ID}};
		\draw[white, fill=white] (-0.0525,0.095) 
		circle [radius=0.007];
	\end{tikzpicture}
	\hspace{-3.0mm}
}
\foreach \x in {A, ..., Z}{\expandafter\xdef\csname orcid\x\endcsname{\noexpand\href{https://orcid.org/\csname orcidauthor\x\endcsname}
		{\noexpand\orcidicon}}
}

\AtBeginDocument{%
	\newwrite\bibnotes
	\def\bibnotesext{Notes.bib}
	\immediate\openout\bibnotes=\jobname\bibnotesext
	\immediate\write\bibnotes{@CONTROL{REVTEX41Control}}
	\immediate\write\bibnotes{@CONTROL{%
			apsrev41Control,author="08",editor="1",pages="1",title="1",year="1"}}
	\if@filesw
	\immediate\write\@auxout{\string\citation{apsrev41Control}}%
	\fi
}%
\begin{document}

\title{Electron-phonon coupling induced topological phase transition in an $\alpha$-$T_{3}$ Haldane-Holstein model}

\author{Mijanur Islam$^\S$\orcidA{}}
\email[]{mislam@iitg.ac.in}

\author{Kuntal Bhattacharyya$^\S$\orcidB{}}
\email[]{kuntalphy@iitg.ac.in}

\author{Saurabh Basu}
\email[]{saurabh@iitg.ac.in}
\affiliation{Department of Physics, Indian Institute of Technology Guwahati, Guwahati-781039, Assam, India}
\begin{abstract}
\noindent We present impelling evidence of topological phase transitions induced by electron-phonon (e-ph) coupling in an $\alpha$-$T_3$ Haldane-Holstein model that facilitates smooth tunability between graphene ($\alpha=0$) and a dice lattice $(\alpha=1)$. The e-ph coupling has been incorporated via the Lang-Firsov transformation which adequately captures the polaron physics in the high frequency (anti-adiabatic) regime, and yields an effective Hamiltonian through zero phonon averaging at $T=0$. While exploring the signature of phase transitions driven by polaron and its interplay with the parameter $\alpha$, we identify two regions based on the values of $\alpha$, namely, the low to intermediate range $(0 < \alpha \le 0.6)$ and larger values of $\alpha~(0.6 < \alpha < 1)$, where the topological transitions host distinct behaviour. There exists a single critical e-ph coupling strength for the former, below which the system behaves as a topological insulator characterized by edge modes, finite Chern number, and Hall conductivity, with all of them vanishing above this value, and the system undergoes a spectral gap closing transition. Further, the critical coupling strength depends upon $\alpha$. For the latter case $(0.6 < \alpha < 1)$, the scenario is more interesting where there are two critical values of the e-ph coupling at which trivial-topological-trivial 
and topological-topological-trivial phase transitions occur. Our study shows a significant difference with regard to the well-known unique transition occurring at $\alpha = 0.5$ (or at $0.7$) in the absence of the e-ph coupling, and thus underscores the importance of interaction effects on topological phase transitions.  
\end{abstract}

\maketitle
\def\thefootnote{\S}\footnotetext{These authors contributed equally to this work.}
\def\thefootnote{\ddag}\footnotetext{The author supervised the work.}

\section{Introduction}
 
Upsurge in generating the topological phases in condensed matter systems has been a modern trend for the last few decades~\cite{Hasan2010,Qi2011}. Although the concept of topology has been prevalent in mathematics for a long time back, it gained enormous attention in modern condensed matter systems thanks to the pioneering work by Thouless~\etal~\cite{Thouless1982}. The TKNN formalism serves as the fundamental tool for understanding the topological nature associated with the quantized plateaus of the quantum Hall (QHE)~\cite{Klitzing1986} systems. In the subsequent years, the interest in predicting new topological phases remains unabated in two~\cite{Kane2005,Bernevig2006} and three~\cite{Moore2007,Fu2007, Konig2007,Hsieh2008} dimensions, topological semimetals~\cite{Yan2017,Armitage2018}, topological superconductors~\cite{Tanaka2012} and many more. Moreover, with the discovery of symmetry-protected topological phases in such systems, a continuous phase transition can be realized between states with same symmetry but different topology, has led to the study of exotic topological materials~\cite{Hasan2010,Qi2011,Kosterlitz1973,Senthil2015}. The topological properties of these phases are protected against disorder. Owing to such robustness to external perturbations, these systems offer potential applications in modern quantum devices, such as quantum computers etc.

Apart from the QHE observed in the presence of an external magnetic field, there have been efforts to realize similar behaviour even in the absence of a magnetic field~\cite{Nagaosa2010,Chang2013,Chang2015,Deng2020}. It was first claimed by Haldane~\cite{Haldane1988} that a complex next-nearest neighbour (NNN) hopping with a phase $\phi$ (also called as the Haldane flux) in a two-dimensional honeycomb lattice breaks the time-reversal symmetry (TRS) of the system and the bands are indexed by a topological invariant, known as the Chern number which is analogous to the quantization of the Hall conductivity. Furthermore, the inclusion of a staggered Samenoff mass ($M$) term in such systems results in the breaking of the sublattice symmetry, which is responsible for opening and closing a band gap at the Dirac points (commonly denoted by the $\textbf{K}$ and $\bf{K^\prime}$ points) in the Brillouin zone (BZ).

\begin{figure}
\includegraphics[width=0.73\linewidth]{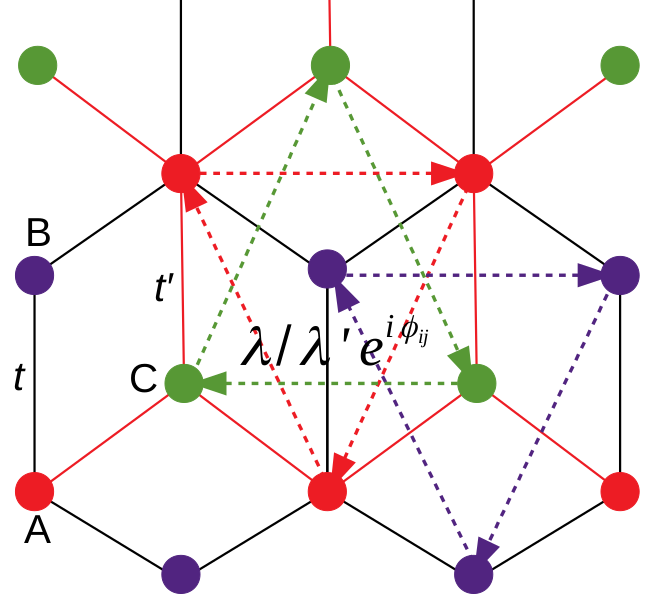}
\caption{The schematic diagram of an $\alpha$-$T_{3}$ lattice is shown, where the red, purple, and green circles represent the sublattices A, B, and C sublattices, respectively. The NN hopping strength between A and B sublattices (solid black line) is $t$, while it is $t^\prime=\alpha t$ between A and C sublattices (solid red line). The NNN hopping between A-B-A (dashed red) or B-A-B (dashed purple) is $\lambda e^{i\phi_{ij}}$, while through C, it is $\lambda^\prime e^{i\phi_{jk}}$ between A-C-A (dashed red) and C-A-C (dashed green). Here $\lambda^\prime=\alpha\lambda$ and the phase $\phi_{ij}$ ($-\phi_{ij}$) denotes the clockwise (anticlockwise) direction.}
\label{fig:model}
\end{figure}

Of late, the study of topology in multiband systems has emerged as one of the fundamental areas that has reshaped the overall scenario of modern condensed matter physics~\cite{Beugeling2015,Hu2023,Huang2022,Zhang2014,Tang2011,Liu2012,Trescher2012,Andrijauskas2015,Jaworowski2015}. With the advent of two-dimensional (2D) graphene-like materials, immense interest has been drawn to studying electronic and transport properties in honeycomb lattices and its variants~\cite{Neto2009,Novoselov2004}. Unlike bare graphene, where the electronic properties are characterized by the Dirac quasiparticles in the low-energy limit, there exists a variant of the honeycomb lattice with $T_{3}$-symmetry, known as the $T_{3}$ or the dice lattice~\cite{Sutherland1986,Vidal1988,Xu2017,Urban2011,BerciouxII2011,Malcolm2016,Vigh2013,Bercioux2009,Wang2011,Dey2020,Mondal2023}, where the low-energy spectrum of the lattice is governed by the Dirac-Weyl quasiparticles. The $T_{3}$ lattice exhibiting pseudospin $S=1$ states can be thought of as an extension to the bare graphene (pseudospin $S=1/2$), fabricated by an additional atom at the centre of the hexagon and may be visualized as comprising of a \text{`C'} sublattice as shown in Fig.\,\ref{fig:model}. 

It has been proposed that the dice lattice can be realized in a cold-atom experimental set-up by three counter-propagating laser beams~\cite{Bercioux2009}. Furthermore, the realization of such lattice is proposed by growing a heterostructure consisting of a trilayer of cubic lattices, namely, SrTiO$_3$/SrIrO$_3$/SrTiO$_3$ in the (111) direction~\cite{Wang2011}. Interstingly, a smooth transformation from pseudospin $S=1/2$ state to pseudospin $S=1$ can be apprehended using a more generalized version of the $T_{3}$ system, known as the $\alpha$-$T_{3}$ lattice~\cite{Raoux2014,Malcolm2015,Illes2015,Wang2021}, where $\alpha$ is the strength of the nearest neighbour (NN) hopping from the central atom (shown by the red line in Fig.\,\ref{fig:model}). Therefore, an $\alpha$-$T_{3}$ lattice sets the two extreme limits, namely, $\alpha=0$ (graphene) and $\alpha=1$ (dice), between which a continuous control is possible by parametrizing $\alpha$ suitably through a variation of the Berry phase, proportional to $\alpha$. Raoux~\etal~\cite{Raoux2014} have explored the role of the Berry phase in $\alpha$-$T_{3}$ lattice and shown that the orbital susceptibility of the system undergoes a transition from a diamagnetic ($\alpha=0$) to a paramagnetic ($\alpha=1$) at a critical $\alpha_{c}=0.495$  while the Berry phase changes from $\pi$ (graphene) to $0$ (dice). Due to the presence of the additional atom at the centre, the low-energy spectrum of the $\alpha$-$T_{3}$ lattice, governed by the tight-binding Dirac-Weyl Hamiltonian, embodies two dispersive bands that are linear in momenta and a dispersionless flat band at zero-energy. Malcolm~\etal~\cite{Malcolm2015} have shown that the bulk dispersion of a quantum well structure made of Hg$_{1-x}$Cd$_x$Te can be linked to the low-lying dispersion of an $\alpha$-$T_{3}$ lattice with an effective $\alpha=1/\sqrt{3}$ at a critical Cd doping concentration, $x_{c}\approx 0.17$. In recent times, such $\alpha$-$T_{3}$ lattices with enlarged spin states ($S>1/2$) have offered a series of studies on equilibrium~\cite{Raoux2014,Malcolm2015,Illes2015,WangJJ2020,Wang2021,Kovacs2017,Islam2017,Illes2017,Biswas2016, Biswas2018,SinghI2023,Gorbar2019,Chen2019,Illes2016,ChenLei2019,Roslyak2021,Balassis2020,Wang2020,Han2022,SunJ2022,Iurov2022,Iurov2020,SinghII2023,IslamM2023,IslamM2022} and nonequilibrium~\cite{Dey2018,Dey2019,Mojarro2020,Niu2022,Tamang2021,Tamang2023,Iurov2019} properties. To highlight a few, people have investigated the role of Berry phase~\cite{Raoux2014,Illes2015,Dey2018,Iurov2019,SinghI2023}, valley-polarized transport~\cite{Islam2017,Niu2022} that may be applicable to valleytronics, Klein tunneling~\cite{Illes2017}, optical conductivity~\cite{Illes2015,Illes2016,Kovacs2017,Chen2019,ChenLei2019,Han2022}, magnetotransport properties, such as Shubnikov–de Hass oscillation and quantized Hall conductivity~\cite{Biswas2016,Illes2015,WangJJ2020,SinghI2023}, floquet dynamics~\cite{Dey2019,Tamang2021}, and other topological signatures~\cite{Mondal2023,BerciouxII2011,Wang2021} in $\alpha$-$T_{3}$ lattices. As there was evidence in the past which demonstrated the topology of the multiband systems, such as,  kagomé~\cite{Tang2011,Liu2012,Trescher2012}, Lieb~\cite{Jaworowski2015}, and dice~\cite{Xu2017,Vigh2013,Wang2011,BerciouxII2011,Dey2020,Mondal2023} lattices could be turned into a Chern insulator by tuning the system parameters, recently, Dey~\etal~\cite{Dey2019} have attempted to show the topological transition at a critical $\alpha_{c}=1/\sqrt{2}$ via irradiating an $\alpha$-$T_{3}$ lattice with a circularly polarized light. 
 
However, the role of many-body correlations like electron-electron and electron-phonon interactions in inducing the topological phase transition in an $\alpha$-$T_{3}$ lattice is left unnoticed.
To the best of our knowledge, the previous proposals of topological phase transition are mainly in non-interacting systems (single-particle picture) where the topology is solely described by the properties of the bands of the electron~\cite{Bradlyn2017,Mondal2023,BerciouxII2011,Wang2021} or by other external means~\cite{Dey2019,Chen2019}. Nevertheless, attempts were made to investigate the topological phase transitions driven by many-body interactions in the past~\cite{Rachel2018,Sorensen2005,Raghu2008,Hohenadler2013,Wang2014,Sun2009,Varney2010,Dong2018,Repellin2017,Sergi2020}, most of which are devoted to explaining the effects of electronic correlations on the topological phases of matter. On the contrary, the role of electron-phonon (e-ph) interaction in such contexts has been scarce. Historically, the e-ph interaction has been proven to deliver promising discoveries in solids~\cite{Devreese2009,Alexandrov2010,Chatterjee2017} starting from the inducing superconductivity~\cite{Tinkham2004,Frohlich1950,Bardeen1951}, transport in three-dimensional materials~\cite{Ziman2001}, low-dimensional polaronic effects~\cite{Yan2013,Samsonidze2007,Mukhopadhyay1995,Mukhopadhyay1999,Challis2003}, Peierls transition~\cite{Kivelson1983,Senna1993,Luo2023,Zhang2023}, charge density wave~\cite{Xie2022,Campetella2023,Luo2022} formation in solids to the Fermi-polarons in ultracold gases~\cite{Koschorreck2012,Kohstall2012,Schirotzek2009,Scazza2017}, topological signatures in novel systems~\cite{Garate2013,Li2013,Li2015,Heid2017} etc. More recently, Bose polaron~\cite{Mostaan2023,Schmidt2022,Hu2016,Jorgensen2016}, phonon-induced Floquet topological phases~\cite{Chaudhary2020,Hubener2018} and several others have been actively explored.  In a polar or an ionic crystal, a propagating electron distorts the lattice structure. Consequently, a net polarization potential emerges due to the interaction between the electron and the \text{`oscillating'} lattice because of which the electron itself may get trapped. The quasiparticles generated because of this interaction can be identified as electrons dressed with phonon clouds, which are commonly known as polarons. Depending on the strength of the e-ph interaction, the polarons can be self-trapped (strong coupling limit) or delocalized (weak coupling limit). In a tight binding system, the electron is found to be strongly bound to its own lattice site, and that electron can participate in forming the polaron by interacting with the onsite phonons. Therefore, the radius of the polaron in such narrow-band systems is short-ranged and does not spread over many lattice sites. The polarons in such systems are often called as the small polarons or the Holstein polarons~\cite{Holstein1959,Bonca1999}.  The polaron formation in tight binding systems can be realized through an interaction between the \text{`extra'} fermionic impurity and the phonons in the system. We shall include the \text{`polaron'} physics in a non-trivially gapped system. As discussed earlier, breaking the TRS via complex NNN hopping is a starting point for our study. Specifically, we assume that this \text{`impurity'} moving in the $\alpha$-$T_{3}$ lattice and interacting with the lattice vibrations, gives rise to the non-trivial spectral gap arises by polarons formed in a Haldane Chern insulator. 

The main focus of this paper is whether e-ph interaction in such a Haldane-Holstein model on an $\alpha$-$T_{3}$ lattice can induce topological phase transitions. If yes, whether these transitions are accompanied by the conventional wisdom, such as (dis)appearance of conducting edge modes, abrupt change in the topological invariant, the behaviour of the anomalous Hall conductivity etc. This interaction-driven topology may provide a favourable platform to explore exotic phenomena in the topological materials. It also serves to connect the correlated phenomena in physics with topology. There are a few studies which describe the importance of e-ph coupling in determining the nontrivial phases in the Haldane Chern insulator~\cite{Cangemi2019,Camacho2019}, graphene nanoribbons~\cite{Calvo2018}, and in other two-dimensional materials~\cite{Sergi2020,Pimenov2021,Calvo2018,Medina2022,Lu2023}. Cangemi~\etal~\cite{Cangemi2019} have proposed a topological quantum transition in a Haldane Chern insulator driven by the e-ph coupling, where they have shown the system undergoes a nontrivial to trivial transition with increasing e-ph coupling strength,  where the average number of fermions shows a sharp discontinuity at the transition point indicating a topological transition. Along the same line, Camacho~\etal~\cite{Camacho2019} have calculated a phonon-induced transverse Hall effect through the \text{``composite Berry phase''} and shown the conductance jumps from zero to a finite value accompanied by a nonzero Chern number. Using a diagrammatic technique in the continuum Dirac model, Pimenov~\etal~\cite{Pimenov2021} have reported a similar observation as in Ref.~\cite{Camacho2019}. These studies largely encourage looking for systems that exhibit a nontrivial phase and possible phase transitions upon suitably tuning the strength of the e-ph coupling. 

However, there is hardly any study revealing the effects of e-ph interaction for an $\alpha$-$T_{3}$ system in the presence of a topological gap. Therefore, in this study, we aim to explore the role of e-ph coupling in stimulating the nontrivial topological phases in the $\alpha$-$T_{3}$ lattice, which may provide a fruitful prescription to understand the interaction-driven topology in other novel systems.

The remainder of this paper is organized in the following manner. In Sec. \ref{Sec:model}, we describe our system and present the model Hamiltonian of a polaronic $\alpha$-$T_{3}$ lattice, which is written in Sec. \ref{textmodel} under the framework of the Haldane model modified by a Holstein term accounting for the e-ph coupling. In Sec. \ref{textLFT}, we show the polaron formation in our system employing the Lang-Firsov technique, which works well for the high-frequency (anti-adiabatic) optical phonons. Sec. \ref{textmomentum} deals with the momentum space representation of the model Hamiltonian, which we shall use to calculate the band spectra and topological quantities. Sec. \ref{topological transition} is devoted to studying the topological phase transition driven by polarons, where we present our numerical results of the bulk and edge spectra in Sec. \ref{textbulk} and Sec. \ref{textedge}, respectively, while interpolating between graphene and a dice structure. In these sections, we discuss how the bulk bands behave and, consequently, the appearance of the edge state and their vanishing below and above a particular critical e-ph coupling strength as a function $\alpha$. The results will imply a plausible occurrence of topological phase transitions. In Sec. \ref{textBerryChern}, we confirm the topological transitions induced by polaron via numerically computing the polaronic Chern number and the Berry curvature. The results show the discontinuous jumps in the Chern number diagram at critical values of the e-ph coupling that depend on $\alpha$. Further proof, such as the quantized Hall plateaus below the critical value of e-ph coupling obtained in Sec. \ref{textHall} also signifies the topological transition driven by e-ph interaction in our system. Finally, in Sec. \ref{Sec:summary}, we conclude our results and briefly summarize our findings.              
\label{textintro}   

\section{$\alpha$-$T_{3}$ lattice with electron-phonon interaction}\label{Sec:model}

The $\alpha$-$T_{3}$ lattice is schematically shown in Fig.\,\ref{fig:model}. Each hexagon of the lattice constituting A (red), B (purple), and C (green) lattice site forms the unit cell of an $\alpha$-$T_{3}$ lattice where A and B atoms construct the regular honeycomb (graphene) lattice with NN hopping strength $t$ and C is considered to be the additional atom placed at the centre of each hexagon connected only to the A atoms via a hopping strength $\alpha t$ ($\alpha\leq 1$). The hopping between a C and a B atom is prohibited. We introduce our model Hamiltonian below in the presence of an e-ph interaction.    
\subsection{Haldane-Holstein model for $\alpha$-$T_{3}$ lattice}
We formulate our system in the spirit of a tight-binding Haldane-Holstein Hamiltonian, which is written as 
\begin{eqnarray}
\mathcal{H}&=&\biggl[-t\sum_{\langle i,j\rangle}c^{\dagger}_{i}c_{j}-\alpha t\sum_{\langle j,k \rangle}c^{\dagger}_{j}c_{k}-\frac{\lambda}{3\sqrt{3}}\sum_{\langle\langle i,j\rangle\rangle}e^{i\phi_{ij}}c^{\dagger}_{i}c_{j}\nonumber\\
&&~-\frac{\alpha\lambda}{3\sqrt{3}}\sum_{\langle\langle j,k\rangle\rangle}e^{i\phi_{jk}}c^{\dagger}_{j}c_{k}+H.C.\biggr]+M\sum_{i}c^{\dagger}_{i}S_{z} c_{i}\nonumber\\
&&~+\hbar\omega_{0}\biggl[\sum_{i}\biggl(b^{\dagger}_{i}b_{i}+\frac{1}{2}\biggr)+\lambda_{eph}\sum_{i}c^{\dagger}_{i}c_{i}(b^{\dagger}_{i}+b_{i})\biggr],
\label{Ham:model}
\end{eqnarray}
where $c^{\dagger}_{i,j,k} (c_{i,j,k})$ denotes the electronic creation (annihilation) operator corresponding to A, B, and C sites with $i$, $j$, and $k$ indices, respectively. The first term represents the NN hopping between the A and B sites with hopping amplitude $t$, while the second one stands for that between the A and C sites with a different amplitude $t^\prime=\alpha t$, which is present due to the C atoms of a typical $\alpha$-$T_{3}$ lattice. We denote the NN terms by a single angular bracket, $\langle ... \rangle$. The third term is the Haldane term designated for the next nearest-neighbour (NNN) complex hoppings (denoted by the double angular bracket $\langle\langle... \rangle\rangle$) between A-B-A or B-A-B with an amplitude $\lambda$ and a phase $\phi_{ij}$, where it is $\phi_{ij}$ ($-\phi_{ij}$) when the motion of the electron is clockwise (anticlockwise). The effect of the C-atoms in the NNN A-C-A and C-A-C hoppings is represented by the fourth term with a different strength, $\lambda^\prime=\alpha\lambda$. Therefore, the two limiting cases of our study are the results for graphene $(\alpha=0)$ and dice $(\alpha=1)$ lattices. The fifth term of Eq.\,\eqref{Ham:model} is the Samenoff mass term, $M$ is the mass, and $S_{z}$ is the $z$-component of the pseudospin-1 matrix. The effects of phonon modes are incorporated in the sixth and seventh terms, where the sixth term is the total onsite energy of the phonons denoted by the phononic creation (annihilation) operators, $b^{\dagger}_{i} (b_{i})$ of site $i$ and the last term of this modified Haldane model is the Holstein term that describes the onsite coupling between electrons and the longitudinal optical (LO) phonons with a coupling strength $\lambda_{eph}$, $\hbar\omega_{0}$ being the energy scale of phonons with a dispersionless LO frequency, $\omega_{0}$.
\label{textmodel}

\subsection{Polaronic Hamiltonian: Lang-Firsov approach}
The quasiparticles formed by the interaction between a bosonic lattice field (a phonon) and a fermionic charge carrier (an electron) undergo emission and absorption of virtual phonons by the electrons at $T = 0$. Owing to such an interaction, a net polarization potential is generated in which the electrons may get trapped. These quasiparticles dressed with virtual phonon clouds are known as polarons. For a tight-binding system, the size of a polaron is usually less compared to the lattice constant and is known as a small Holstein polaron. Here, we have only considered the onsite e-ph interaction, and neglecting the interactions of electrons with the NN and NNN site phonons, these being weak enough. To study the effects of the e-ph coupling, we first employ the much celebrated Lang-Firsov transformation (LFT), namely
\begin{equation}
    \tilde{\mathcal{H}}=e^{R}\mathcal{H}e^{-R},
\label{LFT}
\end{equation}
 where the generator of the transformation is given by~\cite{Lang1963} 
 \begin{equation}
     R=\lambda_{eph}\sum_{i}c^{\dagger}_{i}c_{i}(b^{\dagger}_{i}-b_{i}).
 \label{generator}    
 \end{equation} 
This is a coherent transformation of a displaced harmonic oscillator that eliminates the phonon degrees of freedom and transforms the Hamiltonian into that for an effective electronic system. We must specify that this unitary transformation works well in the high-frequency (non-adiabatic) regime, meaning the LO frequency of the phonons is much larger than the other electronic parameters of the system, i.e., when $\omega_{0}\gg t,t^\prime, \lambda, \lambda^\prime, M$ and $\lambda_{eph}$. The LFT transforms the total Hamiltonian \eqref{Ham:model} as (see Appendix \ref{Appendix} for the derivation)
\begin{widetext}
\bea
\tilde{\mathcal{H}}&=&-t\biggl[\sum_{\langle i,j\rangle}c^{\dagger}_{i}c_{j}e^{[X_{i}-X_{j}]}+\alpha \sum_{\langle j,k \rangle}c^{\dagger}_{j}c_{k}e^{[X_{j}-X_{k}]}\biggr]-\frac{\lambda}{3\sqrt{3}}\biggl[\sum_{\langle\langle i,j\rangle\rangle}e^{i\phi_{ij}}c^{\dagger}_{i}c_{j}e^{[X_{i}-X_{j}]}+\alpha\sum_{\langle\langle j,k\rangle\rangle}e^{i\phi_{jk}}c^{\dagger}_{j}c_{k}e^{[X_{j}-X_{k}]}\biggr]
\nonumber\\
\nonumber\\
&&~~~~~~~~~~~~~~~~~~~~~~~~~~~~~~~~~~~~~~~~~~~~~~~~~~~~~~~~~~~~~~~~~~~~~~~~~~+\sum_{i}c^{\dagger}_{i}(MS_{z}-\lambda_{eph}^2 \hbar\omega_{0}I_{3})c_{i}+\hbar\omega_{0}\sum_{i}b^{\dagger}_{i}b_{i},
\label{Ham: mod model}
\eea
\end{widetext}
where $I_{3}$ is a $3\times3$ identity matrix. 

The $X$-terms in the exponent contain the phonon operators as
\begin{equation}
    X_{i}=\lambda_{eph}(b^{\dagger}_{i}-b_{i}).
\end{equation}
At this stage, to eliminate the phonon degrees of freedom, one can obtain a zero-phonon average (at $T=0$), which reads for the exponents as
\begin{eqnarray}
\langle 0\vert e^{[X_{i}-X_{j}]}\vert 0\rangle &=& e^{-\lambda_{eph}^2},
\label{Zero phonon}
\end{eqnarray}
The quantity in RHS of Eq.~\eqref{Zero phonon} is known as the Holstein reduction factor which causes the band narrowing. The last term in Eq.~\eqref{Ham: mod model} becomes zero after zero-phonon averaging. Therefore, in the transformed Hamiltonian \eqref{Ham: mod model}, all the parameters are modified by the e-ph coupling and the effective Hamiltonian becomes   
\begin{eqnarray}
\tilde{\mathcal{H}}_{eff}&=&\langle 0\vert \tilde{\mathcal{H}}\vert 0\rangle=-\tilde{t}~\biggl[\sum_{\langle i,j\rangle}c^{\dagger}_{i}c_{j}+\alpha \sum_{\langle j,k \rangle}c^{\dagger}_{j}c_{k}\biggr]\nonumber\\
&&~~-\frac{\tilde{\lambda}}{3\sqrt{3}}\biggl[\sum_{\langle\langle i,j\rangle\rangle}e^{i\phi_{ij}}c^{\dagger}_{i}c_{j}+\alpha\sum_{\langle\langle j,k\rangle\rangle}e^{i\phi_{jk}}c^{\dagger}_{j}c_{k}\biggr]\nonumber\\
&&~~+\sum_{i}c^{\dagger}_{i}(MS_{z}-\lambda_{eph}^2\hbar\omega_{0}I_{3})c_{i},
\label{Ham: eff model}
\end{eqnarray}
where the reduced Holstein and Haldane amplitudes are renormalized as 
\begin{eqnarray}
\tilde{t}=te^{-\lambda_{eph}^2},~~~~~\tilde{\lambda}=\lambda e^{-\lambda_{eph}^2}.
\label{Holstein amp}
\end{eqnarray}
It is clear from Eq.~\eqref{Ham: eff model} that the signatures of polaron in our system are captured through $\tilde{t}$ and $\tilde{\lambda}$ (both contain $\lambda_{eph}$). As e-ph interaction modifies system parameters, it will be interesting to see how polaron induces a topological phase transition at certain critical e-ph coupling strength. To investigate the same, we need to transform the Hamiltonian \eqref{Ham: eff model} to the momentum ($\bf k$) space and calculate the band structures along with the relevant topological properties.
\label{textLFT}

\subsection{The continuum $\alpha$-$T_{3}$-Holstein Hamiltonian}
The modified $\bf k$-space version of an $\alpha$-$T_{3}$ lattice in the presence of e-ph interaction can be obtained by Fourier transforming the effective Haldane-Holstein Hamiltonian \eqref{Ham: eff model} in a tri-atomic sublattice basis as

\begin{eqnarray}
\tilde{\mathcal{H}}(\bf{k})&=&-\tilde{t}~(h_xS_x+h_yS_y)-\frac{2\tilde{\lambda}\mathfrak{Im}(f_\textbf{k})}{3\sqrt{3}~\cos\varphi}S_{zH}\nonumber\\
&&~~~~~~~~~~~~~+MS_{z}-\lambda_{eph}^2\hbar\omega_{0}I_{3},
\label{Ham:momentum space}
\end{eqnarray}
with 
\begin{widetext}
\bea
S_x&=&\nu \begin{pmatrix}
0 & $cos$\varphi & 0\\
$cos$\varphi & 0 & $sin$\varphi\\
0 & $sin$\varphi & 0
\end{pmatrix},
S_y=-i \begin{pmatrix}
0 & $cos$\varphi & 0\\
-$cos$\varphi & 0 & $sin$\varphi\\
0 & -$sin$\varphi & 0
\end{pmatrix},
S_{z}=\begin{pmatrix}
1 & 0 & 0\\
0 & 0 & 0\\
0 & 0 & -1 
\end{pmatrix},
S_{zH}=\begin{pmatrix}
-$cos$\varphi & 0 & 0\\
0 & $cos$\varphi- $sin$\varphi & 0\\
0 & 0 & $sin$\varphi 
\end{pmatrix},
\label{Sx}
\eea
\end{widetext}
where $S_{x}$ and $S_{y}$ are the $x$ and $y$ components of the pseudospin-1 matrix written in terms of an angle $\varphi$ which is related to the parameter $\alpha$ as $\varphi=\tan^{-1}\alpha$. Specifically, $S_{zH}$ arises due to the presence of the NNN Haldane term. The parameter $\nu = \pm 1$ denote the valleys $\textbf{K}$ and $\bf{K^\prime}$ located at $\textbf{K}=(4\pi/3\sqrt 3a,0)$ and ${\bf{K^\prime}} = (-4\pi/3\sqrt 3a,0)$. The polaronic contributions to Eq. \eqref{Ham:momentum space} enter through $\tilde{t}$, $\tilde{\lambda}$ (defined in Eq. \eqref{Holstein amp}) and the last term in Eq. \eqref{Ham:momentum space}. $h_x$, $h_y$ and $f_{\bf_{k}}$ in Eq. \eqref{Ham:momentum space} are given as

\begin{equation}
h_x= \sum_{i=1}^{3} {\cos(\bf{k.d_{i}})},~~h_y= \sum_{i=1}^{3} \sin(\bf{k}.\bf{d_{i}}),
\label{h}
\end{equation}
\begin{eqnarray}
f_{\bf_{k}}&=&\sum_{i=1}^{3} e^{(i\bf{k}.\bf{a_{i}})},
\label{f}
\end{eqnarray}
where the coordinates of the NN sites are ${\bf{d_{1}}}=(\sqrt{3}a/2,a/2)$, ${\bf{d_{2}}}=(-\sqrt{3}a/2,a/2)$ and ${\bf{d_{3}}}=(0,-a)$, while that of the NNN sites are ${\bf{a_{1}}}=(\sqrt{3}a/2,3a/2)$, ${\bf{a_{2}}}=(-\sqrt{3}a/2,3a/2)$ and ${\bf{a_{3}}}=(\sqrt{3}a,0)$, $a$ being the lattice constant. Henceforth, we shall use the $\textbf{k}$-space Haldane-Holstein Hamiltonian in Eq. \eqref{Ham:momentum space} extensively for the rest of the paper.   

To obtain the low-energy limit of the above Bloch Hamiltonian, we must expand Eq. \eqref{Ham:momentum space} in the vicinity of the Dirac points around $\textbf{K}$ and $\bf{K^\prime}$ valleys and linearize it which takes the form of a pseudospin-1 Dirac-Weyl Hamiltonian for the polaronic $\alpha$-$T_{3}$ lattice as

\begin{widetext}
\bea
\tilde{\mathcal{H}}(\bf{q})&=&
\hbar \tilde{v_{f}}\begin{pmatrix}
M-\lambda_{eph}^2\hbar\omega_{0} & (\nu q_{x}-iq_{y})~$cos$\varphi & 0\\
(\nu q_{x}+iq_{y})~$cos$\varphi & -\lambda_{eph}^2\hbar\omega_{0} & (\nu q_{x}-iq_{y})~$sin$\varphi\\
0 & (\nu q_{x}+iq_{y})~$sin$\varphi & -M-\lambda_{eph}^2\hbar\omega_{0}
\end{pmatrix}-\frac{\tilde{\lambda}\nu}{\cos\varphi}\begin{pmatrix}
-$cos$\varphi & 0 & 0\\
0 & $cos$\varphi- $sin$\varphi & 0\\
0 & 0 & $sin$\varphi 
\end{pmatrix},
\label{Ham: low-energy}
\eea
\end{widetext}
with $\hbar\tilde{v_{f}}=3a\tilde{t}/2\cos\varphi$ and ${\bf{q}}=(q_{x},q_{y})=\bf{k}-\textbf{K}$ or ($\bf{k}-\bf{K^\prime}$). 

It is well known that the Dirac-Weyl Hamiltonian \eqref{Ham: low-energy} represents two dispersive bands, namely the valance band (VB) and the conduction band (CB), along with a dispersionless flat band (FB) for graphene ($\alpha=0$) and dice ($\alpha=1$) lattices, and a distorted FB for $0<\alpha<1$~\cite{Wang2021}. In our case, all of these are modified by the polaronic factors through $\tilde{t}$ and $\tilde{\lambda}$ defined in Eq. \eqref{Holstein amp}.
\label{textmomentum}

\section{Polaron induced topological features in an $\alpha$-$T_{3}$ lattice}\label{topological transition}
In this section, we present the numerical results of our system and study the effects of e-ph interaction in the context of topological phase transition.
 
\subsection{Bulk spectral properties}
\begin{figure}
\includegraphics[width=1.025\linewidth]{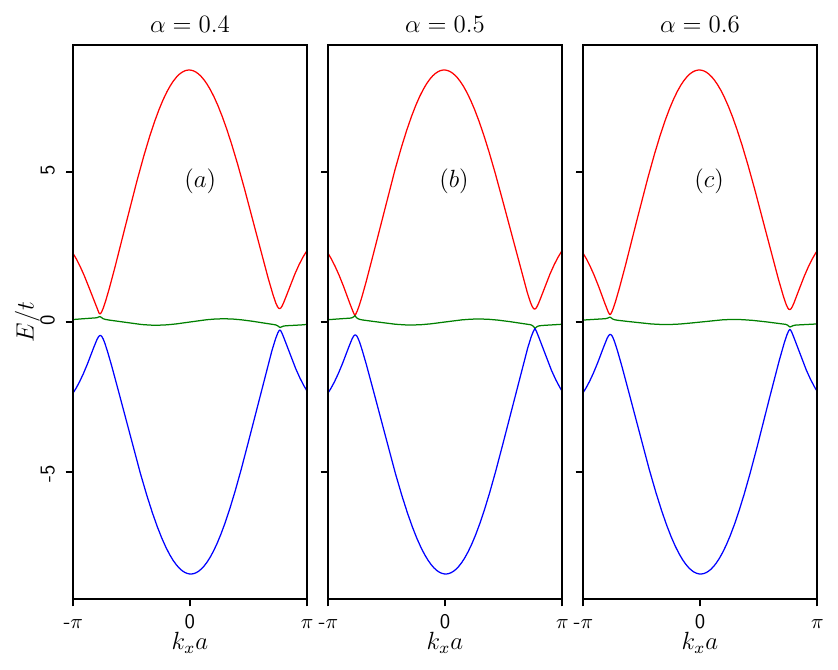}
\caption{The bulk band structures with energy $E$ (in the units of $t$) of the bare Haldane model are shown as a function of dimensionless momenta, $k_x$ (multiplied by the lattice constant) at $k_y=0$ for various values of $\alpha$: $(a)$ $\alpha=0.4$, $(b)$ $\alpha=0.5$, and $(c)$ $\alpha=0.6$. The red, green, and blue colours represent the CB, the FB and the VBs, respectively. Bands are no longer symmetric under the exchange of valleys ($\textbf{K}$ and $\bf{K^\prime}$). The Haldane term is taken as $\lambda=0.1t$.}
\label{fig:bulk_eph0}
\end{figure}

In our study, all the energy parameters are taken in units of $t$, which is set to unity. Further, we fix $a=1$ (lattice constant), $\phi_{ij}=\pi/2$ (the Haldane flux) and $\hbar=1$, for convenience.

Before delving into the specifics of the e-ph interaction, let us briefly explore the bare Haldane $\alpha$-$T_3$ lattice. In the absence of e-ph interaction and the mass term, and solely due to the breaking of the time-reversal symmetry by the Haldane term, the original zero-energy FB may get distorted. Additionally, the electronic band structure experiences valley splitting. Fig.\,\ref{fig:bulk_eph0} illustrates the low-energy bands for various values of $\alpha$ within the first BZ. The red, green, and blue colours denote the CB, the FB and the VBs, respectively. In three-band systems, there can be two distinct band gaps at the Dirac points: the gaps between (i) the CB and the distorted FB ($\Delta^{\textbf{K}/\textbf{K}^\prime}_{cf}$), and (ii) the distorted FB and VBs ($\Delta^{\textbf{K}/\textbf{K}^\prime}_{vf}$) at the $\textbf{K}/\textbf{K}^\prime$ points. The middle band exhibits no dispersion at $\alpha = 0$ (not depicted here), but it gets more dispersive with increasing $\alpha$. We see a mild dispersive nature of the FB at $\alpha = 0.4$ as shown in Fig.\,\ref{fig:bulk_eph0}$(a)$. In Fig.\,\ref{fig:bulk_eph0}$(b)$, where $\alpha = 0.5$, the distorted FB now connects with the VB by closing the gap between them at the $\textbf{K}$ valley, while in the other valley ($\textbf{K}^\prime$), the distorted FB connects to the CB. With further increase in $\alpha$, the gap re-opens, as depicted in Fig.\,\ref{fig:bulk_eph0}$(c)$ for $\alpha = 0.6$. In case of $\alpha = 1$ (not shown here), the spectral gap attains its maximum value, and the distorted FB regains its dispersionless behaviour. Notably, at $\alpha = 0.5$ one finds that, $\Delta^\textbf{K}_{cf} \neq 0$, but $\Delta^{\textbf{K}}_{vf} = 0$, and further $\Delta^{\textbf{K}^\prime}_{cf} = 0$, whereas $\Delta^{\textbf{K}^\prime}_{vf} \neq 0$. These findings precisely correspond with the previously reported results concerning the $\alpha$-$T_3$ lattice ~\cite{Wang2021,Dey2019}. 

\begin{figure}
\includegraphics[width=1.025\linewidth]{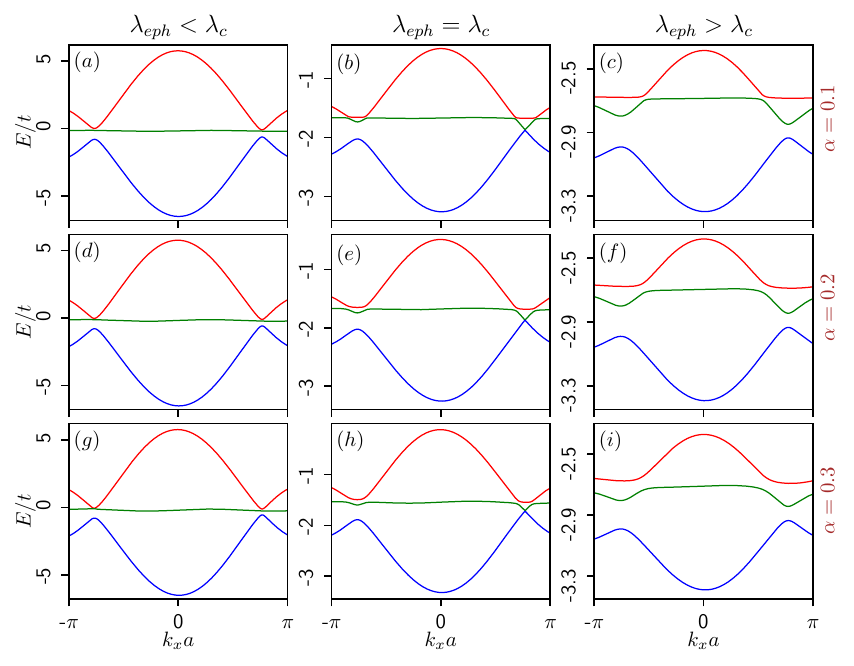}
\caption{Plots of polaronic bulk band structure with energy $E$ (in the units of $t$) for lower $\alpha$ values are shown as a function of dimensionless momenta, $k_x$ (multiplied by the lattice constant) at $k_y=0$. (Left column) The dispersions are plotted in the $\lambda_{eph}<\lambda_c$ regime for $(a)$ $\alpha=0.1$, $(d)$ $\alpha=0.2$, and $(g)$ $\alpha=0.3$, at $\lambda_{eph}=0.3$. (Middle column) Those are plotted at the critical $\lambda_{eph}$ ($=\lambda_c$) for $(b)$ $\alpha=0.1$, $\lambda_c=0.49$, $(e)$ $\alpha=0.2$, $\lambda_c=0.48$, and $(h)$ $\alpha=0.3$, $\lambda_c=0.47$. (Right column) The same are shown in the $\lambda_{eph}>\lambda_c$ regime for $(c)$ $\alpha=0.1$, $(f)$ $\alpha=0.2$, and $(i)$ $\alpha=0.3$, at $\lambda_{eph}=0.6$. The red, green, and blue colours represent the CB, the FB and the VBs, respectively. The parameters are taken as $\lambda=0.1t$ and $M=0.05t$. Further, $t$ and $\lambda$ values are modified as $\tilde{t}$ and $\tilde{\lambda}$ as mentioned in the text. The values of $\lambda_{c}$ are mentioned in Table \ref{tab:table1}.}
\label{fig:bulk}
\end{figure}
\begin{figure}
\includegraphics[width=1.025\linewidth]{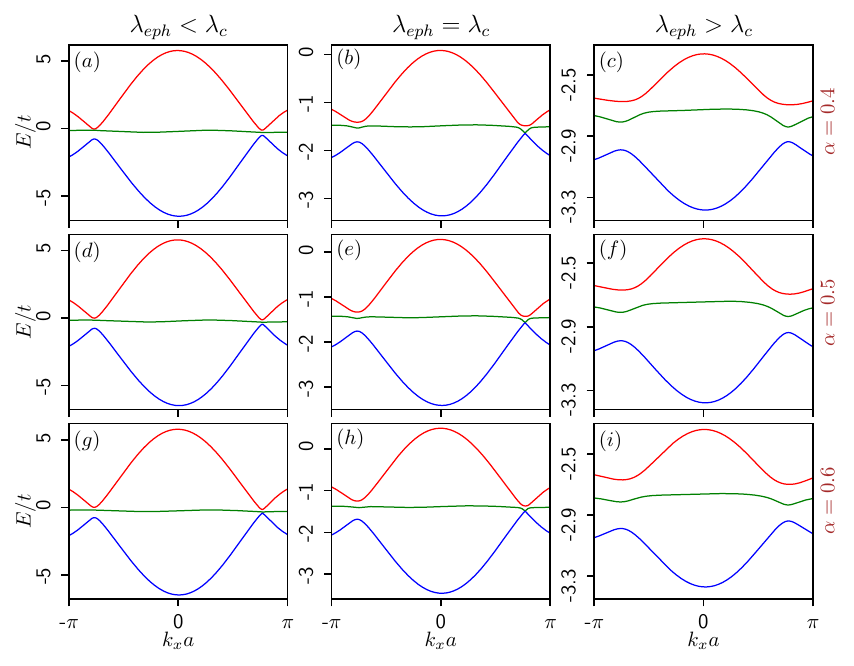}
\caption{Plots of polaronic bulk band structure with energy $E$ (in the units of $t$) for intermediate $\alpha$ values are shown as a function of dimensionless momenta, $k_x$ (multiplied by the lattice constant) at $k_y=0$. (Left column) The dispersions are plotted in the $\lambda_{eph}<\lambda_c$ regime for $(a)$ $\alpha=0.4$, $(d)$ $\alpha=0.5$, and $(g)$ $\alpha=0.6$, at $\lambda_{eph}=0.3$. (Middle column) Those are plotted at the critical $\lambda_{eph}$ ($=\lambda_c$) for $(b)$ $\alpha=0.4$, $\lambda_c=0.46$, $(e)$ $\alpha=0.5$, $\lambda_c=0.45$, and $(h)$ $\alpha=0.6$, $\lambda_c=0.43$. (Right column) The same are shown in the $\lambda_{eph}>\lambda_c$ regime for $(c)$ $\alpha=0.4$, $(f)$ $\alpha=0.5$, and $(i)$ $\alpha=0.6$, at $\lambda_{eph}=0.6$. The red, green, and blue colours represent the CB, the FB and the VBs, respectively. The parameters are taken as $\lambda=0.1t$ and $M=0.05t$. Further, $t$ and $\lambda$ values are modified as $\tilde{t}$ and $\tilde{\lambda}$ as mentioned in the text. The values of $\lambda_{c}$ are mentioned in Table \ref{tab:table1}.}
\label{fig:bulk2}
\end{figure}
\begin{figure}
\includegraphics[width=1.025\linewidth]{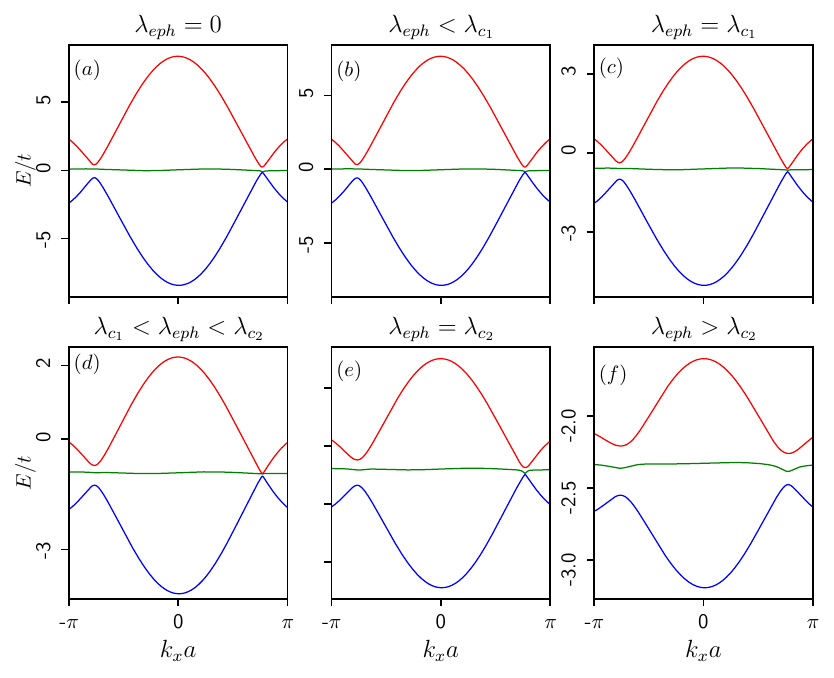}
\caption{Plots of polaronic bulk band structure with energy $E$ (in the units of $t$) for $\alpha=0.7$ are shown as a function of dimensionless momenta, $k_x$ (multiplied by the lattice constant) at $k_y=0$ for $(a)$ $\lambda_{eph} = 0$, $(b)$ $\lambda_{eph}<\lambda_{c_1}$ ($\lambda_{eph}=0.2$), $(c)$ $\lambda_{eph}=\lambda_{c_1}=0.28$, $(d)$ $\lambda_{c_1}<\lambda_{eph}<\lambda_{c_2}$ ($\lambda_{eph}=0.35$), $(e)$ $\lambda_{eph}=\lambda_{c_2}=0.43$, and $(f)$ $\lambda_{eph}>\lambda_{c_2}$ ($\lambda_{eph}=0.6$). The red, green, and blue colours represent the CB, the FB and the VBs, respectively. The parameters are taken as $\lambda=0.1t$ and $M=0.05t$. Further, $t$ and $\lambda$ values are modified as $\tilde{t}$ and $\tilde{\lambda}$ as mentioned in the text. The values of $\lambda_{c_1}$ and $\lambda_{c_2}$ are mentioned in Table \ref{tab:table2}.}
\label{fig:bulk3}
\end{figure}
\begin{figure}
\includegraphics[width=1.025\linewidth]{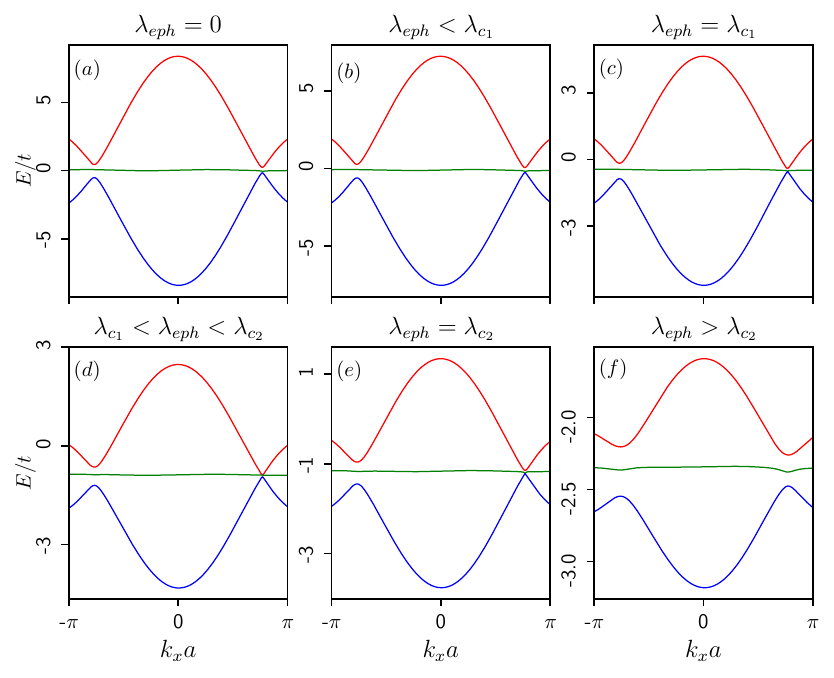}
\caption{Plots of polaronic bulk band structure with energy $E$ (in the units of $t$) for $\alpha=0.8$ are shown as a function of dimensionless momenta, $k_x$ (multiplied by the lattice constant) at $k_y=0$ for $(a)$ $\lambda_{eph} = 0$, $(b)$ $\lambda_{eph}<\lambda_{c_1}$ ($\lambda_{eph}=0.15$), $(c)$ $\lambda_{eph}=\lambda_{c_1}=0.2$, $(d)$ $\lambda_{c_1}<\lambda_{eph}<\lambda_{c_2}$ ($\lambda_{eph}=0.35$), $(e)$ $\lambda_{eph}=\lambda_{c_2}=0.44$, and $(f)$ $\lambda_{eph}>\lambda_{c_2}$ ($\lambda_{eph}=0.6$). The red, green, and blue colours represent the CB, the FB and the VBs, respectively. The parameters are taken as $\lambda=0.1t$ and $M=0.05t$. Further, $t$ and $\lambda$ values are modified as $\tilde{t}$ and $\tilde{\lambda}$ as mentioned in the text. The values of $\lambda_{c_1}$ and $\lambda_{c_2}$ are mentioned in Table \ref{tab:table2}.}
\label{fig:bulk4}
\end{figure}
 Let us include e-ph interaction in the ongoing discussion. We set the mass as $M=0.05t$, NNN hopping as $\lambda=0.1t$ and the phonon-frequency as $\omega_{0}=3t$ which is greater than $t$, $M$, $\lambda$ for non-adiabaticity to be valid. In order to study the topological phases and transitions therein, we first present the bulk spectrum of the $\alpha$-$T_{3}$ system for a few chosen values of $\alpha$ in Fig.\,\ref{fig:bulk} and examine the closing and opening of bulk gaps at the valleys via tuning the e-ph interaction strength $\lambda_{eph}$. As the bulk properties vary with the parameter $\alpha$, we segregate them into two classes of $\alpha$, namely, (i) $0<\alpha\lesssim 0.6$ (from close to the bare graphene to moderate $\alpha$ cases), (ii) $0.6\lesssim\alpha<1$ (from moderate $\alpha$ to Dice lattice). The purpose of such distinction will be clear in a moment. First of all, in Fig.\,\ref{fig:bulk}, we show the bulk energy bands for lower $\alpha$ values, namely, for $\alpha=0.1$, $\alpha=0.2$ and $\alpha=0.3$. As expected, we get three different spectra, namely the VB (shown in blue), the FB (in green), and the CB (in red) as a function of the dimensionless momentum $k_{x}a$ ($k_y$ is set to be zero). The FBs are dispersive (especially for $\alpha>0$ cases) due to the presence of the NNN hopping $\lambda$. Further, we notice a semi-Dirac dispersion, i.e., linear along $k_{y}$ and quadratic along the $k_{x}$ direction. The variations of the bands are shown in three different regimes of $\lambda_{eph}$ i.e. when $\lambda_{eph}<\lambda_{c}$ (left panel), at $\lambda_{eph}=\lambda_{c}$ (middle panel) and then when $\lambda_{eph}>\lambda_{c}$ (right panel), where $\lambda_{c}$ is the critical e-ph coupling strength at which the gap closing ($\Delta^{\textbf{K}}_{vf} = 0$) occurs. These critical points ($\lambda_{c}$) for different $\alpha$ values are listed in Table \ref{tab:table1} and the corresponding plot is shown in the inset $(b)$ of Fig.\,\ref{fig:Chern2}$(i)$.
 
Let us consider the $\alpha=0.1$ case (Figs.\,\ref{fig:bulk}$(a-c)$). As mentioned earlier, the mass term lifts the valley degeneracy. Also, the overall band spectrum is shifted vertically down as we increase $\lambda_{eph}$ further (the FB not being at $E=0$). Interestingly, at the two valleys, $\textbf{K}$ and $\bf{K^\prime}$, the e-ph interaction makes the behaviour of the FBs contradictory (which otherwise looks symmetric when $\lambda_{eph}=0$ (see Fig.\,\ref{fig:bulk_eph0})), especially at $\lambda_{eph}=\lambda_{c}$ points. We clearly notice that in the $\lambda_{eph}<\lambda_{c}$ regime (Fig.\,\ref{fig:bulk}$(a)$), the FB almost touches the CB at both the $\textbf{K}$ and $\bf{K^\prime}$ valleys. Although, a prominent gap between VB and FB is maintained in the $\lambda_{eph}<\lambda_{c}$ regime. However, as soon as $\lambda_{eph}$ reaches a critical value, i.e. when $\lambda_{eph}=\lambda_{c}=0.49$, the FB touches the VB (Fig.\,\ref{fig:bulk}$(b)$) at one of the valleys ($\textbf{K}$) and the band gap closes ($\Delta^{\textbf{K}}_{vf} = 0$), while at the other valley ($\textbf{K}^\prime$), the spectrum remains gapped ($\Delta^{\textbf{K}}_{vf} \neq 0$). The band gap re-opens and the gap persists if we increase $\lambda_{eph}$ further. Beyond $\lambda_{c}$ ($\lambda_{eph}>\lambda_{c}$), the behaviour of the FB is almost similar at both the valleys (Fig.\,\ref{fig:bulk}$(c)$), especially for larger values of $\alpha$. Therefore, both in $\lambda_{eph}<\lambda_{c}$ and $\lambda_{eph}>\lambda_{c}$ regimes, the spectrum remains gapped, implying it to be an insulator, and at $\lambda_{eph}=\lambda_{c}$ the bands touch, signifying a semi-metallic ($SM$) behaviour. 
We need to compute the topological properties for different $\lambda_{eph}$ regimes to confirm the topological nature of the phase, which we shall show in the later sections (\ref{textedge} and \ref{textBerryChern}). This phenomenon of band closing and opening at the Dirac points may give rise to a topological phase transition that is solely caused by tuning the e-ph interaction strength. This is the central result of the paper. Smaller values of $\alpha$ in the range $[0.1:0.3]$ demonstrate similar behaviour (Fig.\,\ref{fig:bulk}) with different $\lambda_{c}$'s (listed in Table \ref{tab:table1}).

The intermediate $\alpha$-cases ($0.4\leq\alpha\leq0.6$) are shown in Fig.\,\ref{fig:bulk2} where we observe the same phenomena, except that one notices for $\alpha=0.6$ case, the FB and the VB nearly touch each other even when $\lambda_{eph}<\lambda_{c}$ (the values of $\lambda_{c}$ are mentioned in Table \ref{tab:table1}) region (can be seen clearly if we zoom in Fig.\,\ref{fig:bulk2}$(g)$). This feature persists for larger values of $\alpha$ ($\alpha > 0.6$) and it needs to be addressed carefully. To do so, we plot the band structure in Fig.\,\ref{fig:bulk3} for $\alpha=0.7$, where it is clearly shown that the VB and the FB touch each other below a certain critical value, namely, $\lambda_{c_1}=0.28$ which may describe a $SM$ phase in the $\lambda_{eph}<\lambda_{c_1}$ regime (Fig.\,\ref{fig:bulk3}$(b)$), and will hold even when $\lambda_{eph}=0$ (Fig.\,\ref{fig:bulk3}$(a)$). Then, in the vicinity of $\lambda_{eph}=\lambda_{c_1}$, the gap between VB and FB opens for the first time (can be seen clearly if we zoom in Fig.\,\ref{fig:bulk3}$(c)$), signalling an insulating behaviour, and the gap stays intact in the $\lambda_{c_1}<\lambda_{eph}<\lambda_{c_2}$ regime (can be seen clearly if we zoom in Fig.\,\ref{fig:bulk3}$(d)$) up to a second critical point, namely, $\lambda_{c_2}=0.43$, at which the gap closes (Fig.\,\ref{fig:bulk3}$(e)$), referring a re-onset of a $SM$ phase. The bulk spectrum is gapped beyond $\lambda_{c_2}$ (Fig.\,\ref{fig:bulk3}$(f)$). 
For $\alpha=0.8$ and $\alpha=0.9$, the scenario is a bit more interesting. Unlike $\alpha=0.7$, for $\alpha=0.8$, we observe in Fig.\,\ref{fig:bulk4}$(a-b)$ that the bulk bands remain gapped (can be seen clearly if we zoom in) in $\lambda_{eph}<\lambda_{c_1}$ regime (including $\lambda_{eph}=0$), signifying an insulating (not $SM$ as for $\alpha=0.7$) phase till $\lambda_{eph}=\lambda_{c_1}=0.2$, where the FB and VB touch each other for the first time (Fig.\,\ref{fig:bulk4}$(c)$) and the insulating to $SM$ transition takes place. As we tune $\lambda_{eph}$ above $\lambda_{c_1}$, we observe the same phenomena as it is shown for $\alpha=0.7$ case (Fig.\,\ref{fig:bulk3}), that is, in $\lambda_{c_1}<\lambda_{eph}<\lambda_{c_2}$ regime (Fig.\,\ref{fig:bulk4}$(d)$), the FB and VB remain gapped (can be seen clearly if we zoom in) denoting an insulating phase till $\lambda_{eph}=\lambda_{c_2}=0.44$ at which the system again shows a $SM$ (Fig.\,\ref{fig:bulk4}$(e)$) nature and for $\lambda_{eph}>\lambda_{c_2}$ (Fig.\,\ref{fig:bulk4}$(f)$) it behaves like an insulator, alike it does for $\alpha=0.7$ case. Similar observations hold for $\alpha=0.9$ (not shown here). The values of $\lambda_{c_1}$ and $\lambda_{c_2}$ for $0.7$, $0.8$, and $0.9$ are listed in Table \ref{tab:table2}. We wish to mention that for $0.8\leq\alpha<1$, this feature of multiple phase transition becomes more prominent as we approach the dice lattice ($\alpha=1$). Therefore, for higher $\alpha$-cases ($\alpha > 0.6$), we encounter two situations, one is for $0.6<\alpha<0.8$ and another is for $0.8\leq\alpha<1$. In the former case, we get two critical $\lambda_{c}$-points, namely $\lambda_{c_1}$ and $\lambda_{c_2}$, below (even when $\lambda_{eph}=0$) and above which the system respectively remains semi-metallic and insulating respectively. In between $\lambda_{c_1}$ and $\lambda_{c_2}$, it behaves like an insulator.   
So, the system undergoes an $SM$-insulator-$SM$-insulator transition in the former case, while in the latter case, the system inherits an insulator-$SM$-insulator-$SM$-insulator transition. 
The nature of the gap (topological or trivial), will be ascertained in Sec. \ref{textedge} and Sec. \ref{textBerryChern}. Hence, the band topology in our study is substantially modified by the polaron formation, which is governed by two factors: the renormalized amplitudes $\tilde{t}$ and $\tilde{\lambda}$ (Eq. \eqref{Holstein amp}) and also the interplay between $MS_z$ and $\lambda_{eph}^2\hbar\omega_0$ (last two terms of Eq. \eqref{Ham:momentum space}). The former causes the band narrowing and the latter is responsible for the competitive effects between the mass term and the polaron shift energy. Moreover, these polaronic markers make the variations of the band spectra (especially those of the FB and the VBs) different for different ranges of $\alpha$. Specifically, for higher values of $\alpha$, the correlation between $M$, $\alpha$ and $\lambda_{eph}$ becomes stronger, giving rise to multiple phase transitions. 
In the case of the dice lattice ($\alpha=1$), the flat band remains flat without any distortion and there is no occurrence of band gap closing phenomena for any values of the e-ph coupling $\lambda_{eph}$ (not shown here). So, no $\lambda_{eph}$ yields a topological phase transition. It is worth mentioning that the values of $\lambda_{c}$ are different for different $\alpha$ cases (see Table \ref{tab:table1} and Table \ref{tab:table2}), which ensures that we shall have a phase transition for all $\alpha$ values between $\alpha=0$ to $1$, albeit with different $\lambda_{c}$ values.

\begin{table}[h!]
    \begin{tabular}{|l|c|r} 
      \hline
      ${\bm{\alpha}}$ & ${\bm{\lambda_{c}}}$\\
      \hline
      0.1 & 0.49 \\
      \hline
      0.2 & 0.48   \\
      \hline
      0.3 & 0.47\\
      \hline
      0.4 & 0.46 \\
      \hline
      0.5 & 0.45   \\
      \hline
      0.6 & 0.43 \\
      \hline
      \end{tabular}
      \caption{Table of $\lambda_{c}$ points for $\alpha$ in the range $0.1<\alpha\leq 0.6$.}
      \label{tab:table1}
      \quad
      
      \begin{tabular}{|l|c|r|} 
      \hline
      ${\bm{\alpha}}$ & ${\bm{\lambda_{c_1}}}$ & ${\bm{\lambda_{c_2}}}$ \\
      \hline
      0.7 & 0.28 & 0.43\\
      \hline
      0.8 & 0.20 & 0.44\\
      \hline
      0.9 &  0.26 & 0.39\\
      \hline
      \end{tabular}
      \caption{Table of $\lambda_{c_1}$ and $\lambda_{c_2}$ points for higher values of $\alpha$.}
      \label{tab:table2}
\end{table}
\label{textbulk} 

\subsection{Edge modes of a semi-infinite $\alpha$-$T_{3}$ ribbon}

In this section, to provide support to the topological properties, we discuss the edge state characteristics of a semi-infinite $\alpha$-$T_{3}$ ribbon in the presence of e-ph coupling. In order to envisage whether the bulk band gap is topologically nontrivial, we inspect the crossings of the edge modes between CB and VB through the FB. The ribbon geometry is considered to exhibit zigzag edges~\cite{Alam2019}. Thus, it is infinite along the $x$-direction, while finite along the $y$-direction, breaking the translational symmetry along one direction ($k_y$ in this case), while the same is protected along the other direction ($k_x$). We have taken the width of the nanoribbon as $N=37$, which satisfies the condition of width $N=3q+1$ ($q$ is an integer), and ensures both the edges are composed of $A$ and $C$ sublattices only. The nontrivial topological signatures are reflected in the edge state spectrum, and the details depend upon the values of $\alpha$. 

\begin{figure}
\includegraphics[width=1.02\linewidth]{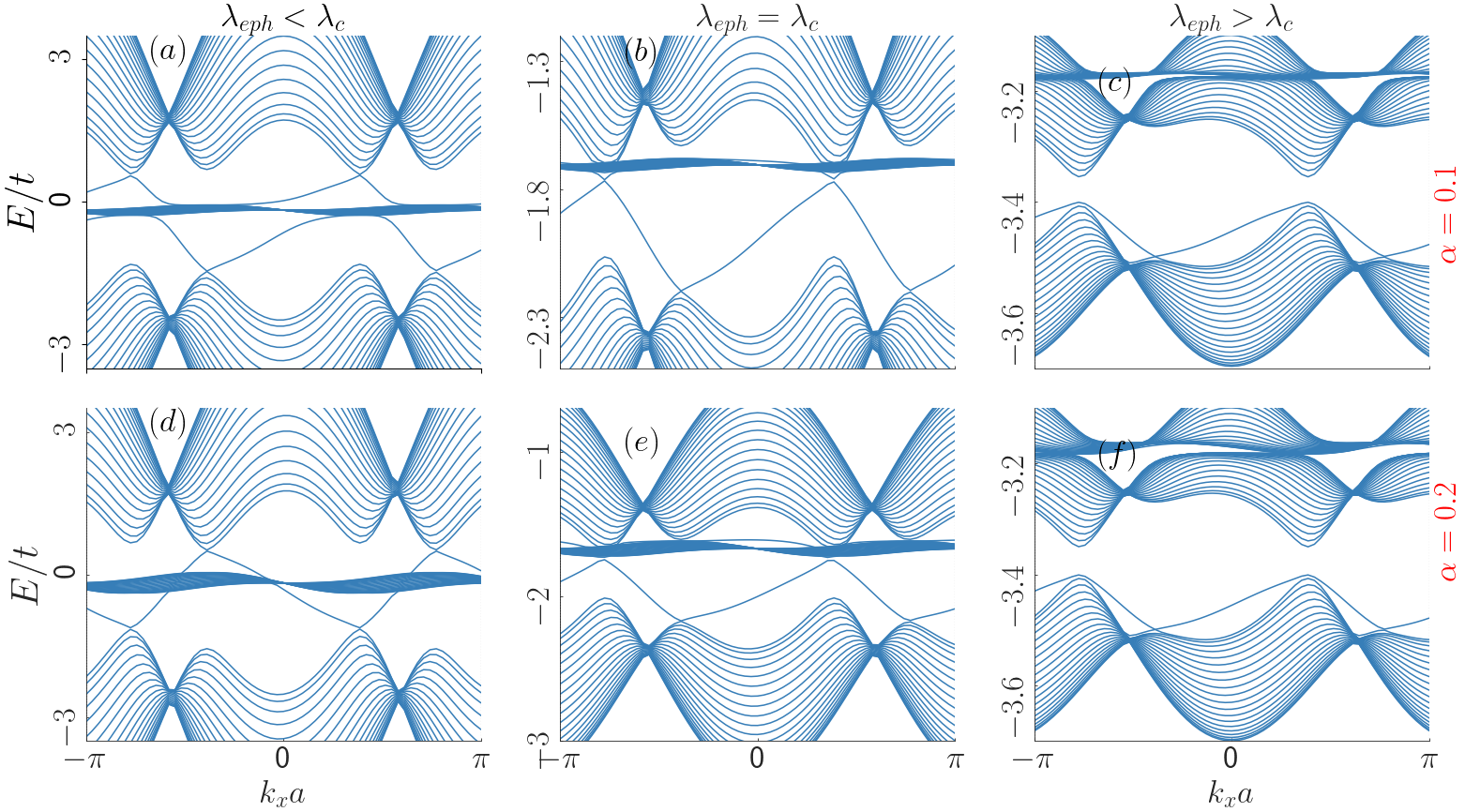}
\caption{Energy spectra (in units of $t$) of the edge states are shown for a zigzag edged semi-infinite ribbon as a function of dimensionless momenta, $k_x$ (multiplied by the lattice constant) of $\alpha=0.1$ for $(a)$ $\lambda_{eph}=0.3$ ($\lambda_{eph}<\lambda_{c}$), $(b)$ $\lambda_{eph}=\lambda_{c}=0.49$, and $(c)$ $\lambda_{eph}=0.6$ ($\lambda_{eph}>\lambda_{c}$), and of $\alpha=0.2$ for $(d)$ $\lambda_{eph}=0.3$ ($\lambda_{eph}<\lambda_{c}$), $(e)$ $\lambda_{eph}=\lambda_{c}=0.48$, and $(f)$ $\lambda_{eph}=0.6$ ($\lambda_{eph}>\lambda_{c}$). Other parameters are the same as those in Fig.\,\ref{fig:bulk}. The values of $\lambda_{c}$ are mentioned in Table \ref{tab:table1}.}
\label{fig:edge}
\end{figure}
We begin by referring to Fig.\,\ref{fig:edge}, where we show the edge states for lower $\alpha$-values ($\alpha=0.1$ and $\alpha=0.2$ marked on the right edge). As stated above in Sec. \ref{textbulk}, the bulk gap closes at a critical $\lambda_{c}$ and it remains gapped corresponding to $\lambda_{eph}<\lambda_{c}$ and $\lambda_{eph}>\lambda_{c}$. We wish to ascertain the existence of edge states that distinguishes a topologically nontrivial phase from a trivial one in both scenarios. Below a critical $\lambda_{c}$, Figs.\,\ref{fig:edge} $(a)$ and $(d)$ display a prominent set of edge states traversing from CB to VB through FB (and vice versa) for $\alpha=0.1$ and $\alpha=0.2$, respectively for the $\lambda_{eph}<\lambda_{c}$ regime. We notice that a pair of edge states emerge from different valleys in the bulk, gather at the FBs and hence cross over to the CBs. By looking at the slope of the edge states, that is, $\partial E/\partial k$, which is a measure of the velocity of the electron, we infer that the flow of the edge currents is counterpropagating, as it should be. These edge states are the chiral edge states of a Chern insulator, appearing in the regime of $\lambda_{eph} < \lambda_{c}$. The nature of the edge states for the $\alpha=0.2$ case are distinct, in the sense that they are crossing the FB at different points. It is also visible in Figs.\,\ref{fig:edge} $(b)$ and $(e)$ that these chiral edges persist up to $\lambda_{eph}=\lambda_{c}$ and disappear beyond that. These are presented in Figs.\,\ref{fig:edge} $(c)$ and $(f)$, that for values above $\lambda_{c}$, the edge states completely disappear and bulk spectra become gapped, signifying the transition of the system to a trivial phase. The critical values of $\lambda_{eph}$ corresponding to the transitions for the $\alpha=0.1$ and $\alpha=0.2$ cases are listed in Table \ref{tab:table1}. Therefore, in $\alpha$-$T_{3}$ systems (with smaller $\alpha$ values), one can generate topological insulating phases via only tuning $\lambda_{eph}$ for a particular value of $\alpha$ below a certain $\lambda_{c}$, beyond which the system goes into a trivial insulating phase.

\begin{figure}
\includegraphics[width=1.02\linewidth]{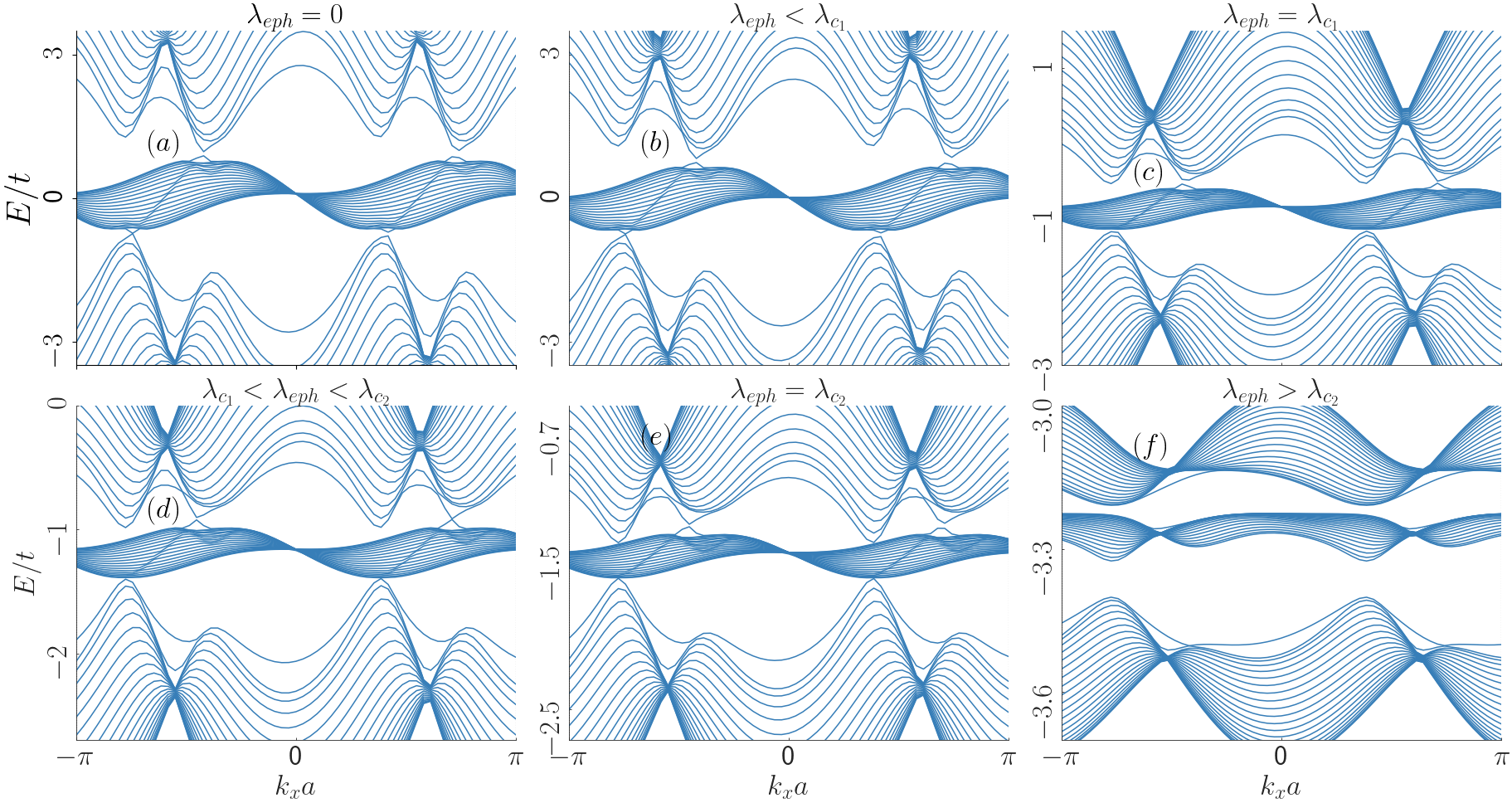}
\caption{Energy spectra (in units of $t$) of the edge states are shown for a zigzag edged semi-infinite ribbon as a function of dimensionless momenta, $k_x$ (multiplied by the lattice constant) of $\alpha=0.7$ for $(a)$ $\lambda_{eph}=0$, $(b)$ $\lambda_{eph}=0.2$ ($\lambda_{eph}<\lambda_{c_1}$), $(c)$ $\lambda_{eph}=\lambda_{c_1}=0.28$, $(d)$ $\lambda_{eph}=0.35$ ($\lambda_{c_1}<\lambda_{eph}<\lambda_{c_2}$), $(e)$ $\lambda_{eph}=\lambda_{c_2}=0.43$, and $(f)$ $\lambda_{eph}=0.6$ ($\lambda_{eph}>\lambda_{c_2}$). Other parameters are the same as those in Fig.\,\ref{fig:bulk3}. The values of $\lambda_{c_1}$ and $\lambda_{c_2}$ are mentioned in Table \ref{tab:table2}.}
\label{fig:edge2}
\end{figure}
\begin{figure}
\includegraphics[width=1.02\linewidth]{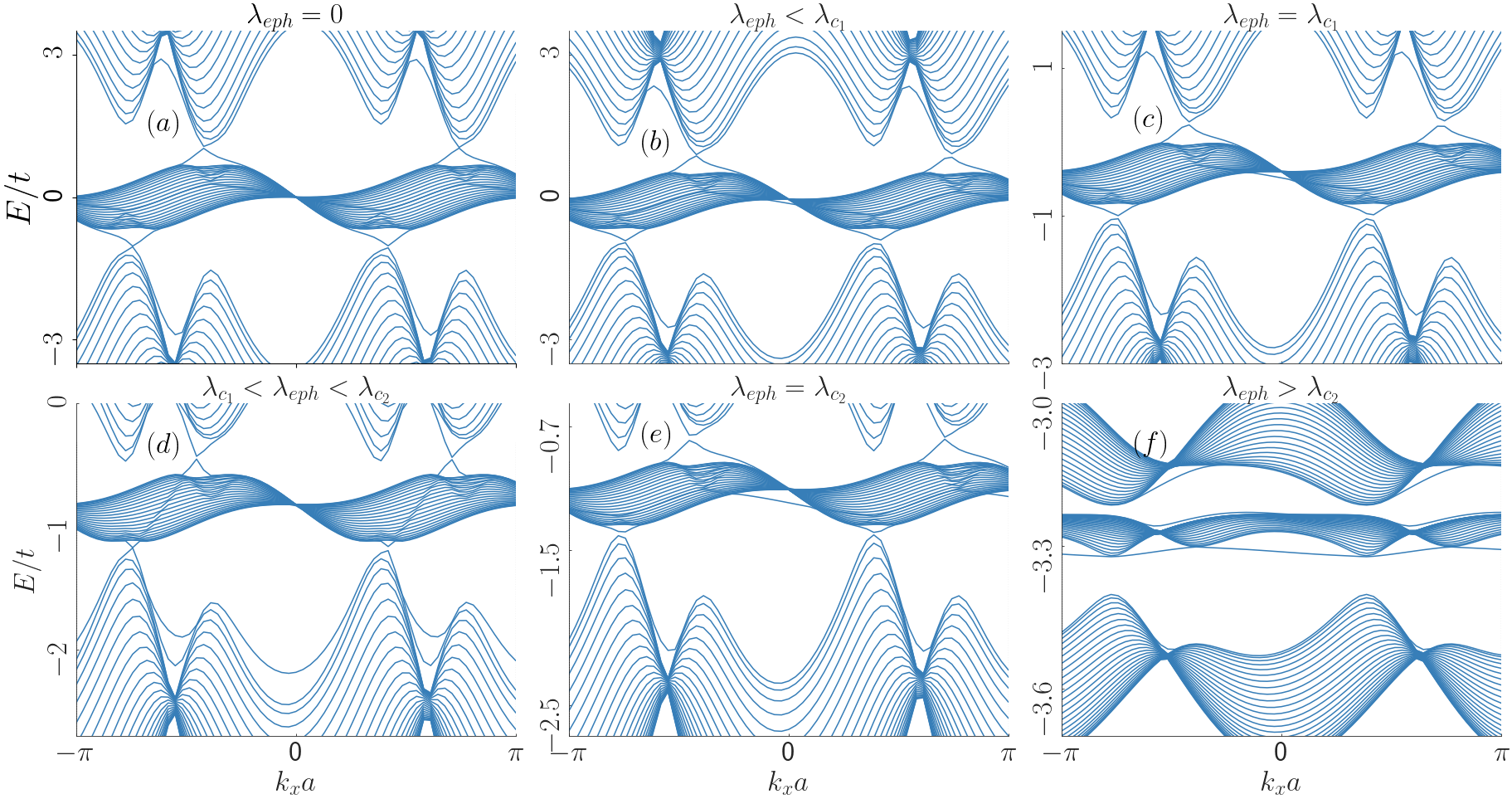}
\caption{Energy spectra (in units of $t$) of the edge states are shown for a zigzag edged semi-infinite ribbon as a function of dimensionless momenta, $k_x$ (multiplied by the lattice constant) of $\alpha=0.8$ for $(a)$ $\lambda_{eph}=0$, $(b)$ $\lambda_{eph}=0.15$ ($\lambda_{eph}<\lambda_{c_1}$), $(c)$ $\lambda_{eph}=\lambda_{c_1}=0.2$, $(d)$ $\lambda_{eph}=0.35$ ($\lambda_{c_1}<\lambda_{eph}<\lambda_{c_2}$), $(e)$ $\lambda_{eph}=\lambda_{c_2}=0.44$, and $(f)$ $\lambda_{eph}=0.6$ ($\lambda_{eph}>\lambda_{c_2}$). Other parameters are the same as those in Fig.\,\ref{fig:bulk4}. The values of $\lambda_{c_1}$ and $\lambda_{c_2}$ are mentioned in Table \ref{tab:table2}.}
\label{fig:edge3}
\end{figure}
Next, let us study the characteristics of the edge states for higher $\alpha$ values, and as a specific case, consider $\alpha=0.7$, presented in Fig.\,\ref{fig:edge2}. In reference to its bulk properties displayed in Fig.\,\ref{fig:bulk3}, we shall examine the edge states for different regimes of $\lambda_{eph}$. As discussed in Figs.\,\ref{fig:bulk3}$(a-b)$, the bulk FB and VB remain in contact with each other for the $\lambda_{eph}\lesssim\lambda_{c_1}$ region,  we notice its signature in Figs.\,\ref{fig:edge2}$(a-b)$, where a pair of counterpropagating edge states emerge near each $\bf{K}$-valley, passing through the FB for the $\lambda_{eph}\lesssim\lambda_{c_1}$ regime. However, in this regime of $\lambda_{eph}$, the notion of edge states is not important as the system does not have any bulk gap, inferring it to be a usual semi-metal. The edge states connecting the VB and CB through the FB are gapped till $\lambda_{eph}=\lambda_{c_1}$ at which the edge states at one $\bf{K}$-valley touch for the first time (can be seen clearly if we zoom in Fig.\,\ref{fig:edge2}$(c)$), thereby generating a conducting edge mode. In the intermediate region (see Fig.\,\ref{fig:edge2}$(d)$), i.e. for $\lambda_{c_1}<\lambda_{eph}<\lambda_{c_2}$, the system clearly exhibits the presence of edge states indicating a topologically nontrivial (Chern insulating) phase. It is evident in Fig.\,\ref{fig:edge2}$(d)$ that the edge states are counterpropagating, and they cross the FB at the two edges for $\lambda_{c_1}<\lambda_{eph}<\lambda_{c_2}$. Therefore, for $\alpha=0.7$, there seems to be a re-entrant mechanism to the $SM$ phase, which may be achieved entirely by tuning the e-ph coupling strength. Around $\lambda_{eph}=\lambda_{c_2}$, the edge states start fading out (shown in Fig.\,\ref{fig:edge2}$(e)$) at one $\bf{K}$-valley and is completely disappear above $\lambda_{c_2}$ (see Fig.\,\ref{fig:edge2}$(f)$). Undoubtedly, the $\lambda_{eph}>\lambda_{c_2}$ region refers to a trivial insulator with no sign of edge states. 
As suggested in the discussion of the bulk spectra (Sec. \ref{textbulk}) that multiple phase transitions (insulator-$SM$-insulator-$SM$-insulator) can occur for $\alpha>0.7$, we explicitly plot the edge states for  $\alpha=0.8$ in Fig.\,\ref{fig:edge3} which ascertains whether the insulating phases are topological. In Fig.\,\ref{fig:edge3}$(a)$ we notice that in the absence of e-ph coupling, a pair of prominent edge states cross the FB at $\bf{K}$ or $\bf{K^\prime}$ valley, signifying a topologically nontrivial Chern insulating phase (unlike for $\alpha=0.7$ where it is $SM$), which remain intact in the $\lambda_{eph}<\lambda_{c_1}$ regime (Fig. \,\ref{fig:edge3}$(b)$) till $\lambda_{eph}=\lambda_{c_1}$. At this value one pair of edge states becomes gapped at one valley, while in the other valley it remains gapless (Fig.\,\ref{fig:edge3}$(c)$). However, such as for the $\alpha>0.7$ case, it is vividly seen in Figs.\,\ref{fig:edge3}$(d-f)$ that the counterpropagating edge states resurface in the $\lambda_{c_1}<\lambda_{eph}<\lambda_{c_2}$ regime, persist up to $\lambda_{eph}=\lambda_{c_2}$ and completely vanish beyond $\lambda_{c_2}$. So, for $\alpha=0.8$ as well, the re-entrant scenario to the $SM$ phase still holds (also true for $\alpha=0.9$, not shown here).
To confirm that the edge modes indeed correspond to a Chern insulating phase, we compute the topological properties and discuss them for each of the regions of $\lambda_{eph}$ (as indicated above) in the following section (Sec. \ref{textBerryChern}).
The phase transition points, namely $\lambda_{c_1}$ and $\lambda_{c_2}$ for $\alpha=0.7,0. 8$ and $0.9$ are listed in Table \ref{tab:table2}.
   
\label{textedge} 
 
\subsection{Berry curvature and Chern number}

To ascertain the topological signatures in the $\alpha$-$T_{3}$ induced by the e-ph coupling, we numerically compute the topological ingredients, namely the (polaronic) Berry curvature and the Chern number. We also obtain the phase diagram containing the Chern number and e-ph coupling strength. In a usual $\alpha$-$T_{3}$ lattice, due to the TRS breaking NNN Haldane term, the system exhibits a nonzero Chern number. The onsite Samenoff mass term that breaks the valley degeneracy also plays a crucial role in band opening at high symmetry Dirac points.
\begin{figure}
\includegraphics[width=1.02\linewidth]{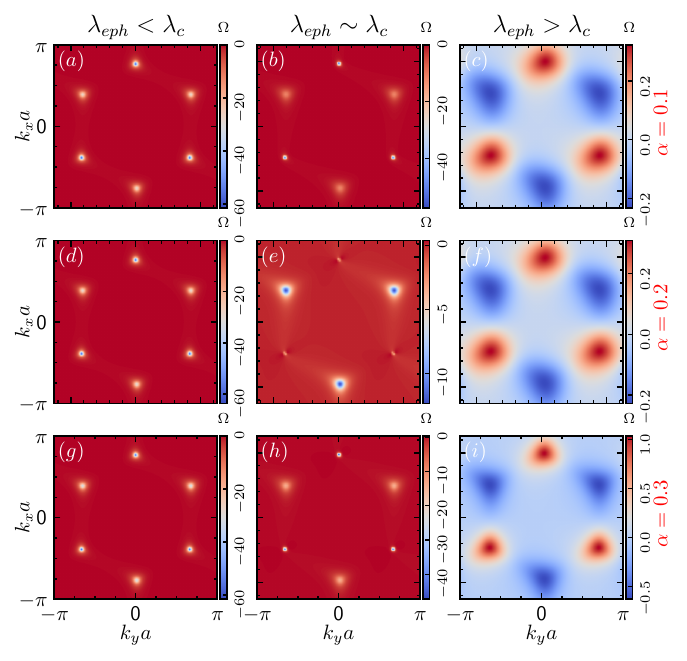}
\caption{The Berry curvature corresponding to the VB is presented for lower $\alpha$-values in different regimes of $\lambda_{eph}$.
(Left column) Those are plotted in the $\lambda_{eph}<\lambda_c$ regime for $(a)$ $\alpha=0.1$, $(d)$ $\alpha=0.2$, and $(g)$ $\alpha=0.3$, at $\lambda_{eph}=0.3$. (Middle column) At the critical $\lambda_{eph}$ ($=\lambda_c$) for $(b)$ $\alpha=0.1$, $\lambda_c=0.49$, $(e)$ $\alpha=0.2$, $\lambda_c=0.48$, and $(h)$ $\alpha=0.3$, $\lambda_c=0.47$. (Right column) The same are shown in the $\lambda_{eph}>\lambda_c$ regime for $(c)$ $\alpha=0.1$, $(f)$ $\alpha=0.2$, and $(i)$ $\alpha=0.3$, at $\lambda_{eph}=0.6$. Other parameters are mentioned in Fig.\,\ref{fig:bulk}. The values of $\lambda_{c}$ are mentioned in Table \ref{tab:table1}.}
\label{fig:Berryp123}
\end{figure}

\begin{figure}
\includegraphics[width=1.02\linewidth]{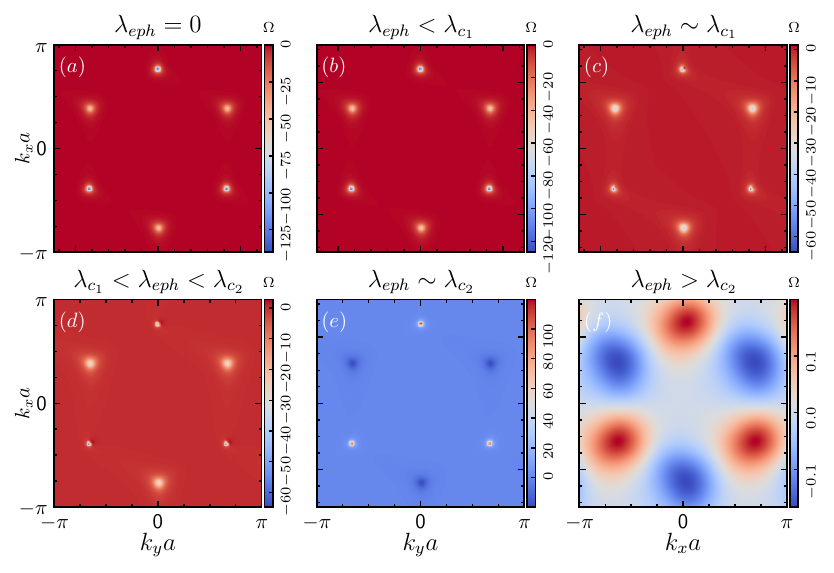}
\caption{The Berry curvature corresponding to the VB is presented for $\alpha = 0.7$ in different regimes of $\lambda_{eph}$ for $(a)$ $\lambda_{eph} = 0$, $(b)$ $\lambda_{eph}<\lambda_{c_1}$ ($\lambda_{eph}=0.2$), $(c)$ $\lambda_{eph}\sim\lambda_{c_1}$ ($\lambda_{eph}=0.26$), $(d)$ $\lambda_{c_1}<\lambda_{eph}<\lambda_{c_2}$ ($\lambda_{eph}=0.35$), $(e)$ $\lambda_{eph}\sim\lambda_{c_2}$ ($\lambda_{eph}=0.41$), and $(f)$ $\lambda_{eph}>\lambda_{c_2}$ ($\lambda_{eph}=0.6$). Other parameters are mentioned in Fig.\,\ref{fig:bulk3}. The values of $\lambda_{c_1}$ and $\lambda_{c_2}$ are mentioned in Table \ref{tab:table2}.}
\label{fig:Berryp7}
\end{figure}
\begin{figure}
\includegraphics[width=1.02\linewidth]{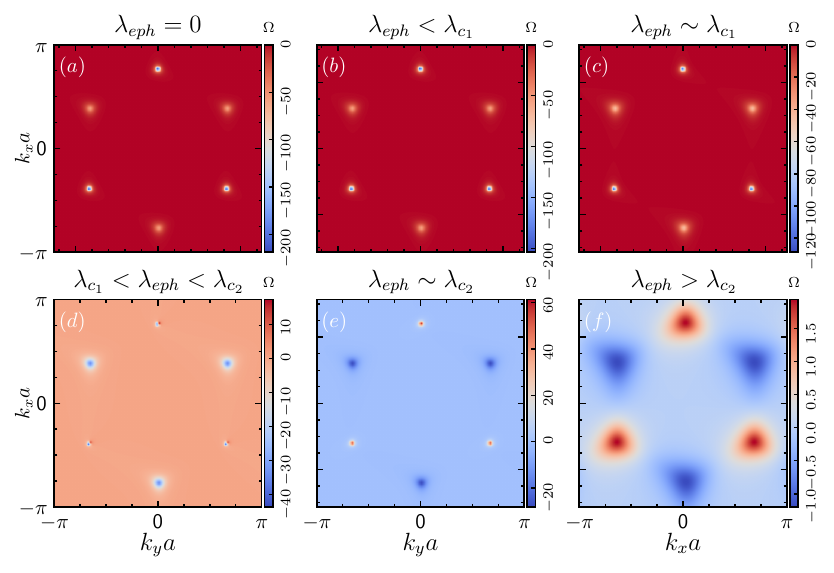}
\caption{The Berry curvature corresponding to the VB is presented for $\alpha = 0.8$ in different regimes of $\lambda_{eph}$ for $(a)$ $\lambda_{eph} = 0$, $(b)$ $\lambda_{eph}<\lambda_{c_1}$ ($\lambda_{eph}=0.15$), $(c)$ $\lambda_{eph}\sim\lambda_{c_1}$ ($\lambda_{eph}=0.18$), $(d)$ $\lambda_{c_1}<\lambda_{eph}<\lambda_{c_2}$ ($\lambda_{eph}=0.35$), $(e)$ $\lambda_{eph}\sim\lambda_{c_2}$ ($\lambda_{eph}=0.42$), and $(f)$ $\lambda_{eph}>\lambda_{c_2}$ ($\lambda_{eph}=0.6$). Other parameters are mentioned in Fig.\,\ref{fig:bulk4}. The values of $\lambda_{c_1}$ and $\lambda_{c_2}$ are mentioned in Table \ref{tab:table2}.}
\label{fig:Berryp8}
\end{figure}
However, our main aim is to investigate how e-ph interaction mediates a nonzero Chern number in the system for a fixed set of other system parameters, namely, $\lambda$ and $M$. We expect that there should exist an interplay between the mass term and the e-ph coupling. Therefore, we may achieve a topological transition only by tuning the strength of e-ph coupling, $\lambda_{eph}$.

\begin{figure}
\includegraphics[width=1.02\linewidth]{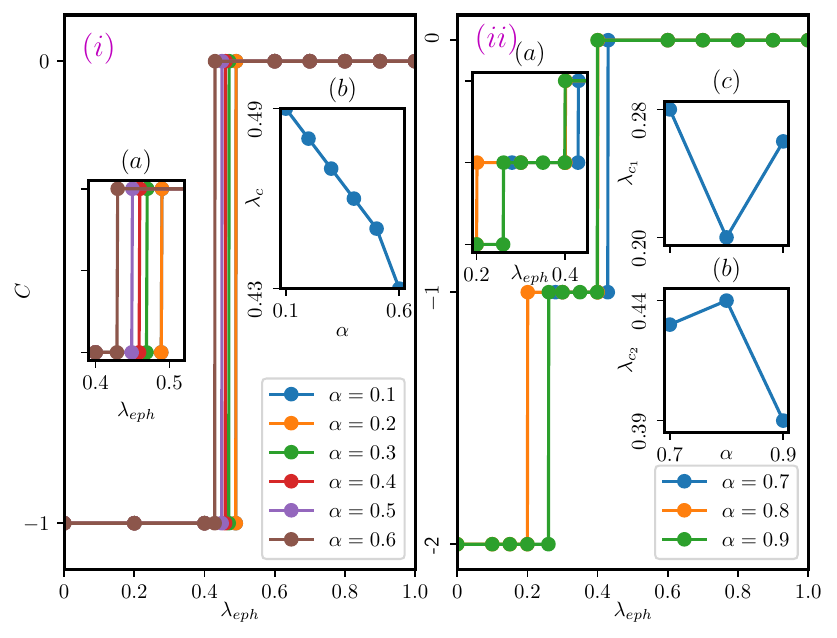}
\caption{The Chern number, $C$ corresponding to the VB as a function of e-ph coupling strength, $\lambda_{eph}$ for $(i)$ lower to intermediate $\alpha$ values ($0< \alpha \leq 0.6$) is shown, while in the inset $(a)$ a zoomed in picture of the transition regions, and in the inset $(b)$, the variation of $\lambda_c$ as a function of $\alpha$ is shown. In $(ii)$, the variations of $C$ for larger $\alpha$ values ($0.6< \alpha \leq 0.9$) are shown. The inset $(a)$ represents a zoomed in picture of the transition regions, while the insets $(b)$ and $(c)$, respectively, display the variations of $\lambda_{c_2}$ and $\lambda_{c_1}$ as a function of $\alpha$. The values of $\lambda_{c}$ are mentioned in Table \ref{tab:table1} and \ref{tab:table2}.}
\label{fig:Chern2}
\end{figure}

The Chern number $(C)$ can be calculated as

\begin{equation}
C=\frac{1}{2\pi}\iint_{BZ} \Omega(k_{x},k_{y})dk_{x}dk_{y},
\label{Chern}
\end{equation}
where $\Omega(k_{x},k_{y})$ is the Berry curvature of our system expressed as
\begin{equation}
\Omega(k_{x},k_{y})=-2i\mathfrak{Im}\biggl[\left\langle\frac{\partial\psi(k_{x},k_{y})}{\partial k_{x}}\bigg|\frac{\partial\psi(k_{x},k_{y})}{\partial k_{y}}\right\rangle\biggr],
\label{Berry}
\end{equation}  
where $\psi(k_{x},k_{y})$ refers to the eigenstate of the modified Haldane model, which is the polaronic bulk band and $\mathfrak{Im}$ denotes the imaginary part. 
The topological phase transition is characterized by a topological invariant. In our study, it is the (polaronic) Chern number, $C$, displayed in Eq.~\eqref{Chern}. In order to calculate so, we first compute the Berry curvature, $\Omega (k_{x},k_{y})$ using Eq.~\eqref{Berry} corresponding to the VB and integrate it over the entire BZ. We plot $\Omega (k_{x},k_{y})$ in Figs.\,\ref{fig:Berryp123}-\ref{fig:Berryp8} and $C$ in Fig.\,\ref{fig:Chern2} to investigate the topological phase transition explicitly mediated through the e-ph coupling.

In Fig.\,\ref{fig:Berryp123}, we show the Berry curvatures in three different regions of $\lambda_{eph}$ i.e., $\lambda_{eph}<\lambda_{c}$, $\lambda_{eph}\sim\lambda_{c}$ (\text{`$\sim$'} sign refers to values close to it, but not at it), and $\lambda_{eph}>\lambda_{c}$ in left, middle and right panels, respectively for smaller values of $\alpha$, namely, $\alpha=0.1$, $\alpha=0.2$ and $\alpha=0.3$ (marked on the right edge). It is generally true that a nonzero Berry curvature is a direct consequence of a nontrivial topology present in the system. The corresponding values seen to be concentrated at the high symmetry points, $\textbf{K}$ and $\bf{K^\prime}$. However, the change in the concentration of the Berry curvatures shown by colourmaps in Figs.\,\ref{fig:Berryp123}-\ref{fig:Berryp8} sets the precursor for any topological transition happening in the system. For $\alpha=0.1$ (Fig.\,\ref{fig:Berryp123}$(a)$), we clearly observe that below the critical $\lambda_{c}$ (i.e. $\lambda_{eph}<\lambda_{c}$ regime) the Berry curvatures are equally distributed in the six corners of the hexagon which defines a topologically nontrivial phase with a nonzero Chern number. But as $\lambda_{eph}$ is increased, the concentration changes. As the Berry curvature is singular at the critical point, we plot it in the vicinity of the critical point ($\lambda_{eph}\sim\lambda_{c}$) shown in Fig.\,\ref{fig:Berryp123}$(b)$. Interestingly, as $\lambda_{eph}$ approaches $\lambda_{c}$, we notice a clear distinction in the concentration of the Berry curvatures at $\textbf{K}$ and $\bf{K^\prime}$ points. At the $\textbf{K}$-point the concentrations are predominantly higher compared to those at the $\bf{K^\prime}$-point. This observation can also be explained via Fig.\,\ref{fig:bulk}$(b)$ where at $\lambda_{eph}=\lambda_{c}$, we see a sharp mismatch in the behaviour of the bulk bands at $\textbf{K}$ and $\bf{K^\prime}$ points, where at one $\textbf{K}$-point, the FB and VB touch each other, while they remain gapped at the $\bf{K^\prime}$-point, displaying the contrasting effects of the e-ph coupling on the FB at two valleys. For other $\alpha$ values ($\alpha=0.2$ and $\alpha=0.3$) that are plotted in Figs.\,\ref{fig:Berryp123} $(e)$ and $(h)$), the distinction between the Berry curvatures at $\textbf{K}$ and $\bf{K^\prime}$-points is much more prominent. In Fig.\,\ref{fig:Berryp123}$(c)$, we show that the Berry curvatures above the critical e-ph coupling strength are almost equal and opposite at $\textbf{K}$ and $\bf{K^\prime}$ points thereby cancelling each other resulting in a topologically trivial phase with a zero Chern number. Hence, till the critical $\lambda_{c}$, the system remains in the topologically nontrivial phase exhibiting a nonzero Chern number. We wish to mention that we have also observed almost similar variations of the Berry curvature by varying the e-ph coupling for the intermediate range of $\alpha$, namely, $\alpha =$ 0.4, 0.5, and 0.6 (not shown here).

To show the variations of the Berry curvature of $\alpha=0.7$, we plot Fig.\,\ref{fig:Berryp7} that can be explained with the help of the bulk and edge spectra displayed in Fig.\,\ref{fig:bulk3} and Fig.\,\ref{fig:edge2}, respectively. Let us first look at Figs.\,\ref{fig:Berryp7} $(a)$ and $(b)$ which is for $\lambda_{eph}=0$ and $\lambda_{eph}<\lambda_{c_1}$, respectively. In this regime, as we have discussed, the spectral gap between the FB and VB vanishes even at $\lambda_{eph}=0$, and remains so till $\lambda_{eph}=\lambda_{c_1}$, manifesting the chiral edge states (see Figs.\,\ref{fig:edge2}$(a-b)$) at the boundaries. However, in this regime, the Berry curvature shows singular behaviour as there is no bulk gap, and consequently, the Chern number is ill-defined. But as we tune $\lambda_{eph}$ further, a bulk gap opens up for the first time at around $\lambda_{eph}\sim\lambda_{c_1}$ (see Fig.\,\ref{fig:bulk3})$(c)$), where we see that the concentrations at the $\textbf{K}$-points start behaving differently than that at $\bf{K^\prime}$-points and are shown in Fig.\,\ref{fig:Berryp7}$(c)$. Beyond $\lambda_{c_1}$, this signature is much more noticeable (can be seen in Fig.\,\ref{fig:Berryp7}$(d)$) and the edge states are prominent (see Fig.\,\ref{fig:edge2}$(d)$) in $\lambda_{c_1}<\lambda_{eph}<\lambda_{c_2}$ regime. In the vicinity of $\lambda_{eph}=\lambda_{c_2}$, the bulk gap closes showing high values for the Berry curvature (see Fig.\,\ref{fig:Berryp7}$(e)$). Finally beyond $\lambda_{c_2}$ (Fig.\,\ref{fig:Berryp7}$(f)$), the variation of the Berry curvature is reminiscent of Figs.\,\ref{fig:Berryp123} $(c)$, $(f)$ and $(i)$ enunciates the onset of a trivial insulating phase. 

The Berry curvature plots for $\alpha=0.8$ are displayed in Fig.\,\ref{fig:Berryp8}. Although the variations in the $\lambda_{eph}<\lambda_{c_1}$ regime (Fig.\,\ref{fig:Berryp8}$(b)$) may look similar to those for $\alpha=0.7$ showing higher values of the Berry curvatures even for $\lambda_{eph}=0$ (Fig.\,\ref{fig:Berryp8}$(a)$), but with the support of the findings of Fig.\,\ref{fig:bulk4} and Fig.\,\ref{fig:edge3} described in Sec. \ref{textbulk} and \ref{textedge} respectively, it is ensured that in the $0\leq\lambda_{eph}<\lambda_{c_1}$ regime, they may correspond to some topological phase (unlike the usual $SM$ phase for $\alpha=0.7$) with conducting edge modes (see Figs.\,\ref{fig:edge3}$(a-b)$) associated with higher Chern numbers. The observations of Figs.\,\ref{fig:Berryp8}$(c-f)$ are almost same as $\alpha=0.7$ case. However, a noticeable dissimilarity with the $\alpha=0.7$-variation in the Berry curvature can be observed for the $\lambda_{c_1}<\lambda_{eph}<\lambda_{c_2}$ regime (Fig.\,\ref{fig:Berryp8}$(d)$), where the disparity between the concentrations of the Berry curvature at two valleys is much more significant compared to that for $\alpha=0.7$. This makes the variation of Fig.\,\ref{fig:Berryp8}$(d)$ distinguishable from Figs.\,\ref{fig:Berryp8}$(a-b)$ denoting different topological phases. 

\begin{figure}
\includegraphics[width=1.02\linewidth]{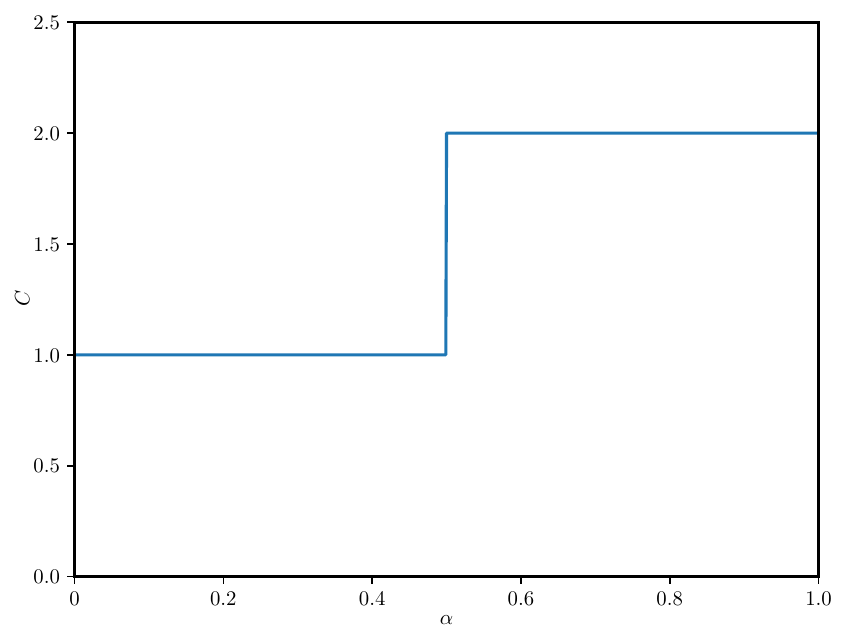}
\caption{Chern number, $C$ as a function of $\alpha$ for the bare (without e-ph coupling and mass term) Haldane model of an $\alpha$-$T_3$ lattice.}
\label{fig:Chern1}
\end{figure}

To confirm the topological phase transition induced by the polaronic interaction in the system, we numerically compute the Chern number, $C$ using Eq.~\eqref{Chern} and examine the variation with the e-ph coupling strength, $\lambda_{eph}$. 

But before going into the intricacies of the electron-phonon interaction, let us take a moment to briefly examine the topological phase transition of the bare Haldane $\alpha$-$T_3$ lattice. In Fig.\ref{fig:Chern1}, we illustrate how a Haldane term on an $\alpha$-$T_3$ lattice renders the system a Chern insulating phase, that is characterized by a nonzero Chern number. Tuning the parameter $\alpha$, a topological phase transition occurs at $\alpha = 0.5$. This transition alters the Chern number of the VB (CB) from $C = -1 (1)$ to a larger Chern number, $C = -2 (2)$. These findings precisely align with previously reported results regarding topological phase transitions in $\alpha$-$T_3$ lattices~\cite{Wang2021}.

Now let us examine the dependency of the Chern number on $\alpha$ in the presence of the e-ph interaction. Fig.\,\ref{fig:Chern2} displays the variations of the Chern number as a function of $\lambda_{eph}$. Here, we display the variations of $C$ separately in two diagrams for lower to intermediate values of $\alpha$ ($\alpha=0.1,0.2, ... ,0.6$) (see Fig.\,\ref{fig:Chern2}$(i)$) and larger values of $\alpha$ ($\alpha=0.7,0.8,0.9$) (see Fig.\,\ref{fig:Chern2}$(ii)$). Starting from the $\alpha=0.1$ case to the intermediate values, such as $\alpha=0.6$, we notice that $C=-1$ up to a critical $\lambda_{c}$, at which $C$ abruptly falls to $C=0$ showing a sharp discontinuity. Therefore, for lower to intermediate cases of $\alpha$ (Fig.\,\ref{fig:Chern2}$(i)$), the system initially behaves like a Chern insulator designated by a nonzero Chern number until the e-ph coupling reaches a certain critical value, $\lambda_{c}$ ($\lambda_{c}$'s are listed in Table \ref{tab:table1}) at which the system undergoes a topological transition accompanied by the closing of the bulk gap and emerging signatures of the edge states. While, beyond $\lambda_{c}$, the system ceases to host edge states which is a typical signature of a trivial insulator for which $C=0$. We display a zoomed in picture of the transition points in the inset $(a)$ of Fig.\,\ref{fig:Chern2}$(i)$ and $(ii)$. In the inset $(b)$ of Fig.\,\ref{fig:Chern2}$(i)$, we depict the variation of the critical $\lambda_c$ with respect to $\alpha$, illustrating a nearly linear decrease with increasing $\alpha$. 

It is understood by now that such variations at higher $\alpha$ values are in contrast to those at lower values of $\alpha$. Let us first consider 
the variation corresponding to $\alpha=0.7$ which is represented via a solid blue line in Fig.\,\ref{fig:Chern2}$(ii)$. Unlike the lower $\alpha$ values, there exist two transition points, namely, $\lambda_{c_1}$ and $\lambda_{c_2}$ for higher $\alpha$ values (listed in Table \ref{tab:table2}). Below the former, the system inherits a conventional $SM$ phase (where $C$ is ill-defined), and above the latter, the system becomes a trivial insulator. Understandably, the $\lambda_{c_1}<\lambda_{eph}<\lambda_{c_2}$ region is our main interest for $\alpha=0.7$, where we find that the Chern number is fixed at $C=-1$ that underscores the emergence of a topologically nontrivial insulating phase, driven entirely by the e-ph coupling. As expected, beyond $\lambda_{c_2}$, $C$ becomes zero confirming the onset of a trivial phase. Thus, the e-ph coupling favours a transition from a semi-metal to a topological insulator, and to a trivial insulator for $\alpha=0.7$.
However, the $0\leq\lambda_{eph}<\lambda_{c_1}$ regime becomes interesting for $\alpha>0.7$ as described earlier in the findings of Figs.\,\ref{fig:edge3} $(a)$ and $(b)$ that the conducting edge modes exist (specifically for $0.8\leq\alpha<1$) in that regime of $\lambda_{eph}$ indicating a topologically nontrivial insulating phase. In Fig.\,\ref{fig:Chern2}$(ii)$, we plot $C$ as a function of $\lambda_{eph}$ for $\alpha=0.8$ denoted by the solid orange line, where higher Chern number, namely $C=-2$ in $0\leq\lambda_{eph}<\lambda_{c_1}$ regime is noted, confirming emergence of a distinct (other than $C=-1$) topological phase. Nevertheless, as we tune $\lambda_{eph}$ further, the scenario becomes exactly the same as $\alpha=0.7$, that is, $C$ changes from $C=-2$ to $C=-1$ at $\lambda_{eph}=\lambda_{c_1}$ signifying a different topological phase that persists in the $\lambda_{c_1}<\lambda_{eph}<\lambda_{c_2}$ regime, which finally vanishes beyond $\lambda_{c_2}$. A similar observation is also shown for $\alpha=0.9$ (marked by the solid green line). Therefore, for the $0.7<\alpha<1$ regime, the system undergoes a transition from one topological phase ($C=-2$) to another ($C=-1$) and hence transits to a trivial ($C=0$) phase, purely mediated all the while by the e-ph coupling. The emergence of $|C|=2$ topological phase in an $\alpha$-$T_{3}$ is a familiar phenomenon obtained by others~\cite{Wang2021,Dey2019,Mondal2023} in the absence of e-ph interaction. For our case, the results completely match with those in Ref.~\cite{Wang2021} corresponding to $\lambda_{eph}\rightarrow 0$ and $M\rightarrow 0$ (shown in Fig.\ref{fig:Chern1}). Moreover, due to the e-ph interaction, we obtain a $|C|=2$ topological phase for lower values of $\lambda_{eph}$, even at $\lambda_{eph}=0$ (in $0\leq\lambda_{eph}<\lambda_{c_1}$ regime) for $0.8\leq\alpha<1$, that is for $\alpha$ values close to the dice lattice limit ($\alpha=1$).    
As earlier, in the inset $(b)$ and $(c)$ of Fig.\,\ref{fig:Chern2}$(ii)$, we show the variation of the critical $\lambda_{c_2}$ and $\lambda_{c_1}$, respectively as a function of $\alpha$. We observe that $\lambda_{c_1}$ initially decreases and then increases with increasing $\alpha$, whereas an opposite trend is observed for $\lambda_{c_2}$, namely, it increases first and hence decreases with increasing $\alpha$. We should mention that the findings of Chern number plots are completely consistent with those of bulk and edge spectra for different regimes of $\lambda_{eph}$. As stated in Sec. \ref{textbulk}, although we have shown a few cases of $\alpha$, it is also important to note that this kind of transition can occur for any value of $\alpha$ ($0< \alpha <1$). Therefore, it seems robust that the polaron formation in $\alpha$-$T_{3}$ lattices induces a topological phase transition generated solely due to the presence of e-ph coupling. 

\begin{figure}
\includegraphics[width=1.02\linewidth]{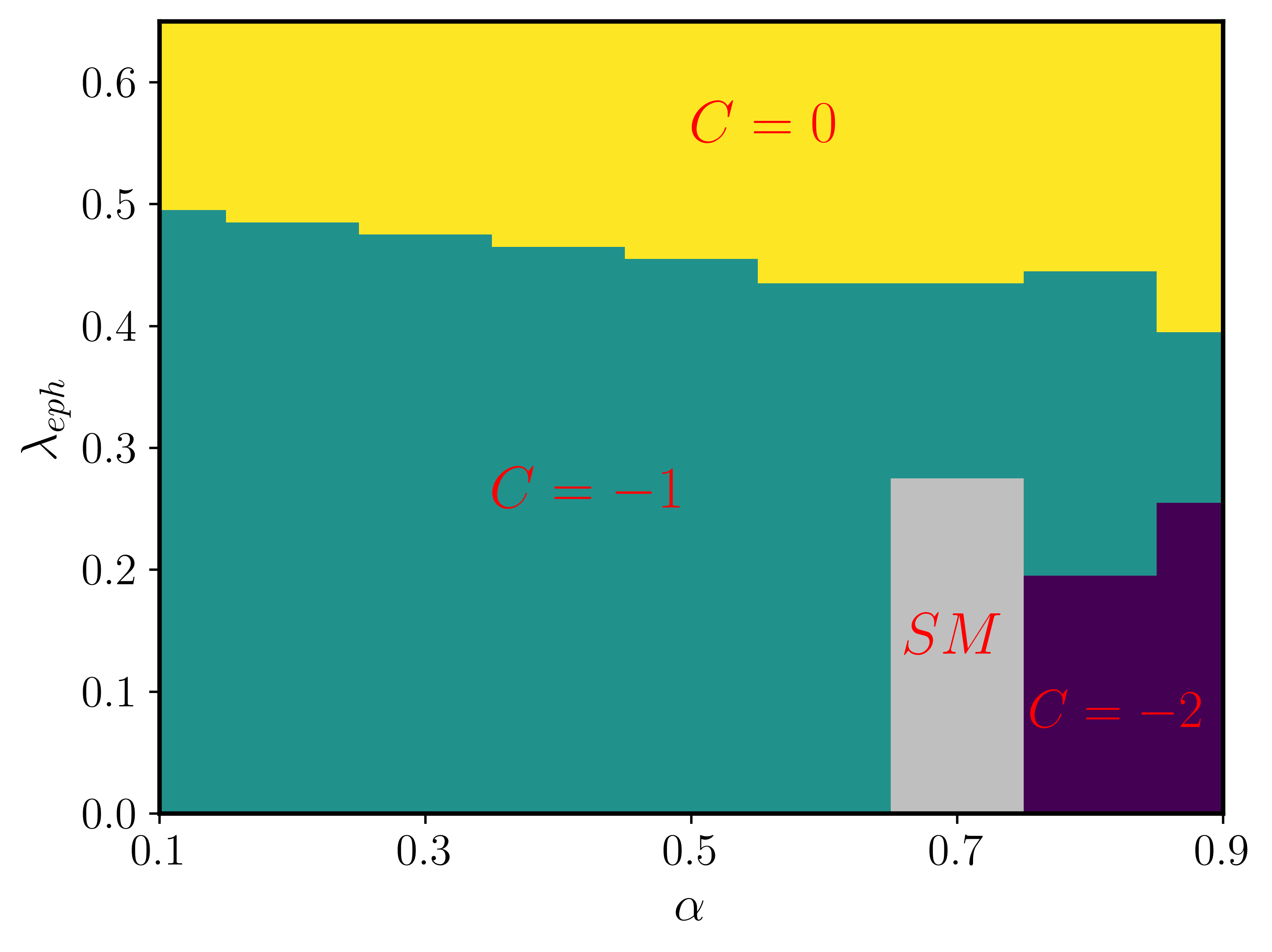}
\caption{The topological phase diagram based on the Chern number ($C$) corresponding to the VB in $\lambda_{eph}-\alpha$ plane. The nonzero $C$ corresponding to the teal region is denoted as $C=-1$, while the yellow region represents the vanishing Chern number ($C=0$). The grey region denotes the $SM$ phase for $0.65\lesssim\alpha\lesssim 0.75$, while the deep purple region stands for a distinct topological phase with $C=-2$ for $0.75\lesssim\alpha< 1$. Other parameters remain the same as mentioned in Fig.\,\ref{fig:bulk}.}
\label{fig:Phase}
\end{figure}

So far we have discussed the topological transitions for different $\alpha$-$T_{3}$ lattices taking discrete values of $\alpha$ in the range $[0:1]$. A phase diagram is hence computed in $\lambda_{eph}-\alpha$ plane to show the exact locations of different (topological/$SM$/trivial) phases in the parameter space. 
The phase diagram containing the Chern number ($C$) corresponding to the VB, in the parameter space defined by e-ph interaction ($\lambda_{eph}$) and $\alpha$ for fixed values of $\lambda$ and $M$, is depicted in Fig.\,\ref{fig:Phase}. 
It is evident that the teal area represents a topological phase of the system with a Chern number as $C=-1$ in the $\lambda_{eph}<\lambda_{c}$ regime for $0<\alpha\lesssim 0.65$ and in the $\lambda_{c_1}\leq\lambda_{eph}\leq\lambda_{c_2}$ regime for $0.65\lesssim\alpha< 1$. Furthermore, for $0.65\lesssim\alpha\lesssim 0.75$ regime, there exists an $SM$ region (where $C$ is ill-defined due to the closing of the bulk band gap) denoted by the grey colour corresponding to $\lambda_{eph}$ values in the $0\leq\lambda_{eph}\leq\lambda_{c_1}$ regime, signifying that the system behaves like a conventional semi-metal. While in the same regime of $\lambda_{eph}$, an $\alpha$-$T_{3}$ lattice with $0.75\lesssim\alpha< 1$ exhibits a distinct topological phase with $C=-2$ (the deep purple region). The yellow region denotes a trivial phase with $C=0$ for all values of $\alpha$ ($0<\alpha<1$) above their respective critical $\lambda_c$-points (listed in Table \ref{tab:table1} and \ref{tab:table2}). It may be noted that varying the parameters $\lambda$ and $M$ can significantly alter the phase diagram. However, we do not show them here.

\label{textBerryChern}

\subsection{Hall conductivity}
\begin{figure}
\includegraphics[width=1.02\linewidth]{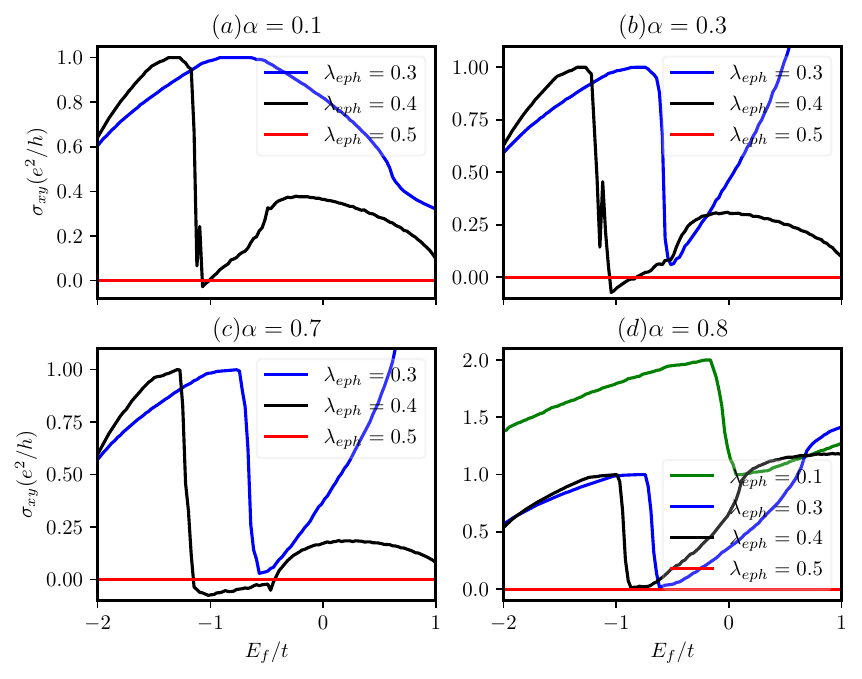}
\caption{The Hall conductivity, $\sigma_{xy}$ as a function of Fermi energy, $E_{f}$ is presented for various values of $\alpha$: $(a)$ $\alpha=0.1$, $(b)$ $\alpha=0.3$, $(c)$ $\alpha=0.7$, and $(d)$ $\alpha=0.8$ for different $\lambda_{eph}$ values that are shown in the inset. Other parameters are the same as mentioned in Fig.\,\ref{fig:bulk}.}
\label{fig:Hall}
\end{figure}

In this section, we numerically compute the polaronic Hall conductivity using the following expression
\begin{equation}
\sigma_{xy}=\frac{e^2}{2\pi h}\sum_{\gamma}\int\frac{dk_{x}dk_{y}}{4\pi^2}f(E_{k_{x},k_{y}}^\gamma)\Omega(k_{x},k_{y}),
\label{Hall}
\end{equation} 
where $e^2/h=\sigma_{0}$ is the scale in which $\sigma_{xy}$ is measured, $E_{k_{x},k_{y}}^\gamma$ is the energy band with the band index $\gamma=-1,0$ and $+1$ corresponding to the VB, FB and the CB, respectively, $f$ denotes the Fermi-Dirac distribution function: $f(E)=[1+e^{(E-E_{F})/k_{B}T}]^{-1}$, $E_{F}$ and $T$ being the Fermi energy and the absolute temperature,  respectively, $\Omega(k_x,k_y)$ being the Berry curvature.  Fig.\,\ref{fig:Hall} display the variations of the polaronic Hall conductivities at $T=0$ as a function of $E_{f}$ for different values of e-ph interaction strength, $\lambda_{eph}$ for $\alpha=0.1$ (see Fig.\,\ref{fig:Hall}$(a)$), $\alpha=0.3$ (Fig.\,\ref{fig:Hall}$(b)$), $\alpha=0.7$ (Fig.\,\ref{fig:Hall}$(c)$), and $\alpha=0.8$ (Fig.\,\ref{fig:Hall}$(d)$). 

As shown in Fig.\,\ref{fig:Hall}$(a)$, the Hall conductivity ($\sigma_{xy}$) is plotted as a function of the Fermi energy ($E_f$) for $\lambda_{eph}=0.3,0.4 $ and $0.5$ marked by solid blue, black and red colours, respectively. It is observed for $\alpha=0.1$, the Hall conductivities initially increase and show tiny plateaus (can also be seen in Fig.\,\ref{fig:Hall}$(b)$ and Fig.\,\ref{fig:Hall}$(c)$ for $\alpha=0.3$ and $\alpha=0.7$, respectively), which are quantized at a value $e^2/h$ for the $\lambda_{eph}<\lambda_c$ regime. In other words, these quantized plateaus occurring at $|C| e^2/h$ (here, $|C|=1$) presented in Fig.\,\ref{fig:Hall}$(a)$ re-confirm that up to a critical e-ph coupling $\lambda_{c}=0.49$, the system behaves like a topological insulator. Beyond the critical $\lambda_{c}$ (denoted by solid red), it becomes a trivial insulator with $\sigma_{xy}=0$ (that is, $C=0$). Similar observation is noted in Fig.\,\ref{fig:Hall}$(c)$ for $\alpha=0.7$  where the plateaus at values $|C| e^2/h$ exist for $\lambda_{eph}$ values that are in the $\lambda_{c_1}<\lambda_{eph}<\lambda_{c_2}$ regime. 

In a scenario where $\lambda_{eph}=0=M$, the Hall conductivity shows plateaus (but with small kinks on the plateaus due to the presence of the distorted FB) as long as the Fermi level lies in between the bulk gap~\cite{Mondal2023}. However, in our case (with $\lambda_{eph}\neq 0$ and $M\neq 0$), the nature of the Hall conductivity deviates significantly as there exists a cumulative effect arising from the interplay of the three parameters, namely the Fermi energy ($E_f$), the mass term ($M$) and the e-ph coupling strength ($\lambda_{eph}$).
The reason for the plateaus to become tinier can be explained with the help of the bulk spectra, which are mainly affected by $M$ and $\lambda_{eph}$ for different $\alpha$ values. It is understandable that a significant width of the plateau is dependent on how accurately we fix the Fermi level in the bulk gap. As discussed in Sec. \ref{textbulk}, the individual bulk bands shrink due to the Holstein factor (Eq. \eqref{Holstein amp}), and the whole band structure shifts vertically down by the polaron shift energy ($\lambda_{eph}^2\hbar\omega_0$). Due to the band narrowing caused by the Holstein factor, the gap between the distorted (because of the Haldane term, $\lambda$) FB and VB decreases, makes it difficult for the Fermi level to lie \textit{\text{`properly'}} in between the bulk gap, making the plateaus less prominent, especially for higher values of $\lambda_{eph}$. Additionally, we find that the increase in $\sigma_{xy}$ as a function of $E_f$ can be explained as follows. As observed in Sec. \ref{textbulk}, $M$ breaks the valley degeneracy and the interplay between $M$ and $\lambda_{eph}$ renders contrasting behaviour of the bulk bands at two valleys, that is, well-gapped at one valley and almost gapless at the other for $\lambda_{eph}<\lambda_c$. As $E_f$ is increased, it is possible that at one valley, $E_f$ may lie well in the gap, while it may lie in the CB as well at the other valley, which will contribute to higher $\sigma_{xy}$. The unusual behaviour of the Hall conductivity due to the presence of a distorted FB has also been reported by Singh~\etal~\cite{SinghI2023} (in the absence of e-ph coupling). Certainly, all of the above discussions become unimportant for $\lambda_{eph}>\lambda_c$. 

Interestingly, for $\alpha=0.8$, the quantized (tiny) Hall plateaus in Fig.\,\ref{fig:Hall}$(d)$ are located at $e^2/h$ (where, $|C|=1$) and $2e^2/h$ (where, $|C|=2$) for $\lambda_{c_1}<\lambda_{eph}<\lambda_{c_2}$ (shown for $\lambda_{eph}=0.3$ and $0.4$, denoted by solid blue and black, respectively) and $0\leq\lambda_{eph}<\lambda_{c_1}$ regimes (shown for $\lambda_{eph}=0.1$, denoted by solid green), respectively, confirm existence of two distinct topological insulating phases with a nonzero $\sigma_{xy}$ (also true for any $\alpha$ in $0.75\lesssim\alpha< 1$ regime), while these plateaus vanish beyond $\lambda_{c_2}$, ascertaining emergence of a trivial insulating phase ($C=0$) with $\sigma_{xy}=0$. Thus, the polaronic Hall conductivity ensures that the system undergoes a transition from a nontrivial insulating phase with quantized plateaus at $|C| e^2/h$ in the $\lambda_{eph}\lesssim\lambda_{c}$ regime to a trivial insulating phase with zero Hall conductivity in the $\lambda_{eph}>\lambda_{c}$ regime.

\label{textHall}

\section{Conclusion}\label{Sec:summary}
To summarize, we have studied the effect of e-ph interaction on inducing a topological phase transition in a Haldane-Holstein model on an $\alpha$-$T_{3}$ lattice. The NN and the complex NNN Haldane hopping amplitudes get renormalized by the Holstein reduction factor showing the signature of polaron formation in the system. The cases of our study are majorly divided into two scenarios, namely lower to intermediate $\alpha$ $(0<\alpha\leq 0.6)$ and higher $(0.6<\alpha< 1)$ values of $\alpha$. With the help of the effective Hamiltonian in $\bf{k}$-space, we have computed the bulk and the edge spectra where it is observed that for the first case, as we increase the e-ph coupling strength $\lambda_{eph}$, the bulk gap between the flat and valance bands closes at a critical coupling strength, namely $\lambda_{c}$ at one $\bf{K}$-valley and re-opens beyond $\lambda_{c}$. This feature explains that the system is characterized by two distinct insulating states below and above $\lambda_{eph}=\lambda_{c}$. Consequently, the conducting edge modes emerge in the $\lambda_{eph}<\lambda_{c}$ regime, which are preserved up to $\lambda_{eph}=\lambda_{c}$, and disappear for $\lambda_{eph}>\lambda_{c}$, signifying a topologically nontrivial to trivial phase transition. In the second case, we encounter a different scenario where the flat and valance bands in the bulk remain gapless for the $0.65\lesssim\alpha\lesssim 0.75$ regime and gapped for $0.75\lesssim\alpha< 1$, till $\lambda_{eph}$ reaches a first critical value, namely $\lambda_{c_1}$ and become gapped till $\lambda_{eph}$ assumes another critical value, namely $\lambda_{c_2}$ where similar gap closing transition takes place. The explicit emergence of conducting edge modes in the $\lambda_{c_1}<\lambda_{eph}<\lambda_{c_2}$ regime both for $0.65\lesssim\alpha\lesssim 0.75$ and $0.75\lesssim\alpha< 1$, and also in $0\leq\lambda_{eph}<\lambda_{c_1}$ regime for $0.75\lesssim\alpha< 1$ that traverse through the FB around the $\bf{K}$ and $\bf{K^\prime}$ valleys makes the latter case more intriguing. It indicates that for $0.65\lesssim\alpha\lesssim 0.75$ $(0.75\lesssim\alpha< 1)$, the system re-enters from a conventional (topological) $SM$ phase (in the $\lambda_{eph}<\lambda_{c_1}$ regime) to a (another) topological one (in the $\lambda_{c_1}<\lambda_{eph}<\lambda_{c_2}$ regime) upon tuning the e-ph coupling strength. The above discussions for both the cases, either with a unique $\lambda_{c}$ or with two $\lambda_{c}$s, namely $\lambda_{c_1}$ and $\lambda_{c_2}$, strongly indicate possibilities of inducing topological phase transition via e-ph coupling in an $\alpha$-$T_{3}$ Haldane-Holstein model. Furthermore, we have numerically computed the Berry curvature and the topological invariant, namely the (polaronic) Chern number ($C$), for different values of $\alpha$. In our study, the evidence of a discontinuous change in $C$ from $|C|=1$ to $|C|=0$ for $0<\alpha\lesssim 0.75$ regime, and from $|C|=2$ to $|C|=1$ and finally to $|C|=0$ for $0.75\lesssim\alpha< 1$ regime exhibiting a jump in the $C$ vs $\lambda_{eph}$ diagram at different critical values of the e-ph coupling for different values of $\alpha$ directly confirms the topological phase transition solely caused by the e-ph interaction, while interpolating $\alpha$ between corresponding lattice structures of graphene to a dice lattice. More specifically, the system under investigation possesses a topological insulating phase accompanied by $|C|=1$ or $|C|=2$ (depending on the range of $\alpha$) below certain critical values of the e-ph coupling strength, and becomes a trivial insulator ($C=0$) above the critical point. We, furthermore, incorporate the above observations in a phase diagram plotted for $C$ in the $\lambda_{eph}-\alpha$ plane. To confirm such phases, and phase transitions from one phase to another, we have calculated the Hall conductivity for a few values of $\alpha$ (both small and large) as a function of $\lambda_{eph}$. The existence (vanishing) of Hall plateaus at $|C| e^2/h$ below (above) a certain critical $\lambda_{c}$ for a particular value of $\alpha$ further substantiates the evidence of topological phase transitions induced by e-ph coupling in our $\alpha$-$T_{3}$ Haldane-Holstein model. We wish to motivate that our study may serve as a powerful tool for understanding the interaction-driven topology in novel quantum systems.\\[6pt] 

\section*{Acknowledgements}
K.B. sincerely thanks IIT Guwahati for providing financial support through the Institute Post Doctoral Fellowship (Ref.~No.~IITG/R$\&$D/IPDF/2023-24/20231003P537). M.I. and K.B. also sincerely acknowledge Ms. Srijata Lahiri and Dr. Sayan Mondal for fruitful discussions.\\[6pt] 

\appendix
\section{\label{app1}Derivation of the modified Haldane-Holstein Hamiltonian for an $\alpha$-$T_{3}$ lattice}
In this section, we briefly derive the major steps to obtain the Hamiltonian (Eq. \eqref{Ham: mod model} of Sec. \ref{textLFT}) modified by the e-ph coupling, employing LFT via the generator of the transformation mentioned in Eq. \eqref{generator}. The transformed Hamiltonian in Eq. \eqref{LFT} can equivalently be expressed by the Baker–Campbell–Hausdorff formula as
\begin{eqnarray}
\tilde{\mathcal{H}}&=&e^{R}\mathcal{H}e^{-R}=\mathcal{H}+[R,\mathcal{H}]+\frac{1}{2!}[R,[R,\mathcal{H}]] \nonumber\\
&&~~~~~~~~~~~~~~~~+\frac{1}{3!}[R,[R,[R,\mathcal{H}]]] + ...~~. 
\label{A1}
\end{eqnarray}
Let us label the terms of Hamiltonian \eqref{Ham:model} as $\mathcal{H}^{(1)}, \mathcal{H}^{(2)}, ... ,$ etc. for the first till the seventh term of Hamiltonian \eqref{Ham:model}, respectively. Now, we transform each individual term as in the following. Combining the NN ($\langle ... \rangle$) terms that are symbolized as $\mathcal{H}^{(1)}$ and $\mathcal{H}^{(2)}$ we can calculate the commutator for $\mathcal{H}^{(12)}\equiv \mathcal{H}^{(1)}+\mathcal{H}^{(2)} $ as
\begin{widetext}
\begin{eqnarray}
[R,\mathcal{H}^{(12)}]&=&\biggl[\lambda_{eph}\sum_{i}c^{\dagger}_{i}c_{i}(b^{\dagger}_{i}-b_{i}),[-t\sum_{\langle i^\prime,j^\prime\rangle}c^{\dagger}_{i^\prime}c_{j^\prime}-\alpha t\sum_{\langle j^\prime,k^\prime \rangle}c^{\dagger}_{j^\prime}c_{k^\prime}]\biggr]\nonumber\\
&&~~~~~~~~~~~~~~~~~~~~~~~~~~~~~~~~~~~=-(1+\alpha)t\sum_{i,\delta}c^{\dagger}_{i}c_{i+\delta}\biggl[\lambda_{eph}[(b^{\dagger}_{i}-b_{i})-(b^{\dagger}_{i+\delta}-b_{i+\delta})]\biggr],
\label{A2}
\end{eqnarray}
and consequently the successive commutators can be obtained as
\begin{equation}
[R,[R,\mathcal{H}^{(12)}]]=-(1+\alpha)t\sum_{i,\delta}c^{\dagger}_{i}c_{i+\delta}[X_{i}-X_{i+\delta}]^2,~ 
[R,[R,[R,\mathcal{H}^{(12)}]]]=-(1+\alpha)t\sum_{i,\delta}c^{\dagger}_{i}c_{i+\delta}[X_{i}-X_{i+\delta}]^3.
\label{A3}
\end{equation} 
where $X_i\equiv \lambda_{eph}(b^{\dagger}_{i}-b_{i})$ and $\delta$ is the NN index, that is, $j=i+\delta$. Therefore, collecting terms in Eq. \eqref{A2} and Eq. \eqref{A3} and using Eq. \eqref{A1} the NN terms are transformed as
\begin{eqnarray}
\tilde{\mathcal{H}}^{(12)}=-(1+\alpha)t\sum_{i,\delta}c^{\dagger}_{i}c_{i+\delta}\biggl[1+[X_{i}-X_{i+\delta}]+\frac{1}{2!}[X_{i}-X_{i+\delta}]^2+\frac{1}{3!}[X_{i}-X_{i+\delta}]^3+ ...\biggr]
\nonumber\\
\nonumber\\
=-(1+\alpha)t\sum_{i,\delta}c^{\dagger}_{i}c_{i+\delta}~e^{[X_{i}-X_{i+\delta}]}.
\label{A4}   
\end{eqnarray}
\end{widetext}
The NNN ($\langle\langle ... \rangle\rangle$) Haldane terms denoted by $\mathcal{H}^{(3)}$ and $\mathcal{H}^{(4)}$ can be transformed in a similar fashion with the NNN index, $\eta$ as 
\begin{equation}
\tilde{\mathcal{H}}^{(34)}=-(1+\alpha)\frac{\lambda}{3\sqrt{3}}\sum_{i,\eta}c^{\dagger}_{i}c_{i+\eta}~e^{[X_{i}-X_{i+\eta}]}e^{i\phi_{i,i+\eta}}, 
\label{ANNN}
\end{equation}
while the onsite mass term, $\mathcal{H}^{(5)} (\equiv\sum_{i}c^{\dagger}_{i}MS_{z} c_{i}$) remains unchanged by the transformation, that is, 
\begin{equation}
\tilde{\mathcal{H}}^{(5)}=\sum_{i}c^{\dagger}_{i}MS_{z} c_{i}. 
\label{AM}
\end{equation}
The phonon energy, $\mathcal{H}^{(6)} (\equiv \hbar\omega_{0}\sum_{i} b^{\dagger}_{i}b_{i}$) and the e-ph interaction term, $\mathcal{H}^{(7)} (\equiv \lambda_{eph}\hbar\omega_{0}\sum_{i}c^{\dagger}_{i}c_{i}(b^{\dagger}_{i}+b_{i})$) can respectively be transformed by Eq. \eqref{A1} as
\begin{equation}
\tilde{\mathcal{H}}^{(6)}=\hbar\omega_{0}\sum_{i}\biggl[b^{\dagger}_{i}b_{i}-\lambda_{eph}c^{\dagger}_{i}c_{i}(b^{\dagger}_{i}+b_{i})+ \lambda_{eph}^2 c^{\dagger}_{i}c_{i}\biggr],
\label{A5}
\end{equation}
\begin{equation}
\tilde{\mathcal{H}}^{(7)}=\hbar\omega_{0}\sum_{i}\biggl[\lambda_{eph}c^{\dagger}_{i}c_{i}(b^{\dagger}_{i}+b_{i})-2\lambda_{eph}^2 c^{\dagger}_{i}c_{i}\biggr], 
\label{A6}
\end{equation}
where in Eqs. \eqref{A5} and \eqref{A6}, we have used the identity, namely $n_i^2=n_i(\equiv c^{\dagger}_{i}c_{i})$ for the fermionic number operator. Hence, summing Eqs. \eqref{A4}-\eqref{A6} we obtain Eq. \eqref{Ham: mod model}.
\label{Appendix}

\bibliography{references}
\end{document}